\documentclass[aps,amsmath,twocolumn,amssymb,floatfix,showpacs,superscriptaddress,nofootinbib,longbibliography]{revtex4-1}
\usepackage{mathtools}
\usepackage{braket}
\usepackage[dvipsnames]{xcolor}
\usepackage{float}
\usepackage{subfigure}
\usepackage[colorlinks=true,linktoc=page,linkcolor=Purple,citecolor=RedViolet,urlcolor=Violet]{hyperref}
\newcommand{\vect}[1]{\boldsymbol{\mathrm{#1}}}
\mathchardef\mhyphen="2D 

\newcommand{\ie}{{i.e.,\,\,}}

\newcommand{\ua}{{\uparrow }}
\newcommand{\da}{{\downarrow }}

\newcommand\bea{\begin{eqnarray}}
\newcommand\eea{\end{eqnarray}}
\newcommand\beq{\begin{equation}}  
\newcommand\eeq{\end{equation}}

\newcommand{\noi}{\noindent}

\newcommand{\non}{\nonumber}  
 
\newcommand{\sgn}{\mathrm{sgn}}
\newcommand{\Pc}{\mathcal{P}}
\newcommand{\mc}{\mathcal}
\newcommand{\A}{\alpha}
\newcommand{\B}{\beta}
\newcommand{\G}{\gamma}
\newcommand{\D}{\delta}

\usepackage[normalem]{ulem}
\definecolor{lime}{HTML}{A6CE39}
\usepackage{sidecap,tikz}
\DeclareRobustCommand{\orcidicon}{\hspace{-1.0mm}
	\begin{tikzpicture}
		\draw[lime, fill=lime] (0.0,0.0) 
		circle [radius=0.15] 
		node[white] {{\fontfamily{qag}\selectfont \tiny \,ID}};
		\draw[white, fill=white] (-0.0525,0.095) 
		circle [radius=0.007];
	\end{tikzpicture}
	\hspace{-3.0mm}
}
\foreach \x in {A, ..., Z}{\expandafter\xdef\csname orcid\x\endcsname{\noexpand\href{https://orcid.org/\csname orcidauthor\x\endcsname}
		{\noexpand\orcidicon}}
}

\AtBeginDocument{%
	\newwrite\bibnotes
	\def\bibnotesext{Notes.bib}
	\immediate\openout\bibnotes=\jobname\bibnotesext
	\immediate\write\bibnotes{@CONTROL{REVTEX41Control}}
	\immediate\write\bibnotes{@CONTROL{%
			apsrev41Control,author="08",editor="1",pages="1",title="1",year="1"}}
	\if@filesw
	\immediate\write\@auxout{\string\citation{apsrev41Control}}%
	\fi
}%
\begin{document}


\title{Distinguishing between topological Majorana and trivial zero modes via transport and shot noise study in an altermagnet heterostructure}

\author{Debashish Mondal\orcidD{}}\thanks{These authors contributed equally to this work.}
\affiliation{Institute of Physics, Sachivalaya Marg, Bhubaneswar-751005, India}
\affiliation{Homi Bhabha National Institute, Training School Complex, Anushakti Nagar, Mumbai 400094, India}

\author{Amartya Pal\orcidA}\thanks{These authors contributed equally to this work.}
\affiliation{Institute of Physics, Sachivalaya Marg, Bhubaneswar-751005, India}
\affiliation{Homi Bhabha National Institute, Training School Complex, Anushakti Nagar, Mumbai 400094, India}

\author{Arijit Saha\orcidC{}}
\email{arijit@iopb.res.in}
\affiliation{Institute of Physics, Sachivalaya Marg, Bhubaneswar-751005, India}
\affiliation{Homi Bhabha National Institute, Training School Complex, Anushakti Nagar, Mumbai 400094, India}

\author{Tanay Nag\orcidB{}}
\email{tanay.nag@hyderabad.bits-pilani.ac.in}
\affiliation{Department of Physics, BITS Pilani-Hydrabad Campus, Telangana 500078, India}

\begin{abstract}
\noi We theoretically investigate the transport and shot noise properties of a one-dimensional semiconducting nanowire with Rashba spin-orbit coupling~(SOC) placed in closed proximity to a bulk 
$s$-wave superconductor and an altermagnet with $d$-wave symmetry. Such heterostructure with vanishing net magnetization manifests itself as an alternative route to anchor Majorana zero modes~(MZMs) characterized by appropriate topological index~(winding number $W$). Interestingly, this system also hosts accidental zero modes~(AZMs) emerged with vanishing topological index indicating their non-topological nature. Furthermore, by incorporating three terminal setup, we explore the transport and shot noise signatures of these zero modes. At zero temperature, we obtain zero bias peak (ZBP) in differential conductance to be quantized with value $|W|\times 2 e^{2}/h$ for MZMs. On the other hand, AZMs exhibit non-quantized value at zero bias. Moreover, zero temperature 
shot noise manifests negative~(positive) value for MZMs~(AZMs) within the bulk gap. At finite temperature, shot noise exhibits negative value~(negative to positive transition) concerning MZMs~(AZMs). Thus, the obtained signatures clearly distinguish between the MZMs and non-topological AZMs. We extend our analysis by switching on the next to nearest neighbor hopping amplitude and SOC. 
Our conclusion remains unaffected for this case as well. Hence, our work paves the way to differentiate between emergent MZMs and AZMs in a semiconductor/ superconductor/ altermagnet heterostructure.
\end{abstract}

\maketitle

\textcolor{blue}{\textit{Introduction.}---} Topological superconductors~(TSC) host Majorana zero modes~(MZMs)~\cite{Kitaev_2001,qi2011topological,Leijnse_2012,Alicea_2012,ramonaquado2017,beenakker2013search}, proposed to be potential candidates for topological quantum computation due to their 
non-Abelian braiding property~\cite{Ivanov2001,freedman2003topological,KITAEV20032,Stern2010,NayakRMP2008}. The idea of TSC was conceived to Kitaev in terms of a one dimensional~(1D) spinless $p$-wave superconductor called Kitaev chain~\cite{Kitaev_2001}. In spite of its intrinsic realization remaining elusive, it motivates researchers to engineer TSC in hybrid setups~\cite{FuPRL2008,LutchynPRL2010,das2012zero,Alicea_2012,Yazdani2013,Yazdani2015,Felix_analytics,Mondal_2023_Shiba}. To realize TSC in heterostructures, there exists an elegant proposal based 
on 1D Rashba nanowire~(NW) placed in close proximity to a bulk $s$-wave superconductor in presence of an external Zeeman field~\cite{LutchynPRL2010,Leijnse_2012,Alicea_2012,Mourik2012Science,das2012zero,ramonaquado2017,Mondal2023_NW,Mondal2024}. Using two-terminal differential conductance and scanning tunneling microscopy~(STM), several transport measurements have been performed in such heterostructure. The emergence of zero bias peak (ZBP) in differential conductance has been reported as an indirect signature of the existence of MZMs in such setup~\cite{Mourik2012Science,das2012zero,Rokhinson2012,Finck2013,Albrecht2016,Deng2016}. 

Note that, basic ingredients to realize TSC are superconductivity, spin-orbit coupling (SOC) and time reversal symmetry~(TRS) breaking element~\cite{Alicea_2012}. For the above heterostructure, the external magnetic filed satisfies the criterion for TRS breaking term. However, it reduces the effective~(proximity induced) superconducting gap. 
To circumvent this issue, very recently new proposals have been put forward based on altermagnet with net vanishing magnetization~\cite{ghorashi2023altermagnetic,Li_PRBL_2024}. 
Altermagnets are special type of collinear antiferromagnetic materials exhibiting momentum dependency in magnetic order parameter~\cite{Smejkal_PRX_1,Smejkal_PRX_2,BhowalPRX2024,Mazin_PRBL_2023,Cheong2024}. In other words, altermagnets can be thought of as $d$-wave Zeeman field, hosting non-relativistic SOC, 
in certain rotationally symmetric materials where anisotropic spin-polarized fermi surface are observed such as $\mathrm{RuO_2}$~\cite{Smejkal_PRX_1}, $\mathrm{MnF_2}$~\cite{BhowalPRX2024}, 
$\mathrm{MnTe}$~\cite{Mazin_PRBL_2023} etc. being the proposed candidate materials. In spite of net zero magnetization, altermagnets lead to spin-splitting of energy bands 
with changing sign in the Brillouin zone in addition to intrinsic broken TRS. 
Hence, they open up a new avenue to engineer Majorana fermions in altermagnet heterostructure~\cite{Smejkal_PRX_1,Smejkal_PRX_2,BhowalPRX2024,Cheong2024,Bai_PRL_2023,Mazin_PRBL_2023,Ouassou_PRL_2023,Papj_PRBL_2023,XZhou2024_PRL,Zhang2024,Li_PRBL_2023,Li_PRBL_2024,
Zhu2023,Li2024,zhu2024field}.

\begin{figure}[]
	\centering
	\subfigure{\includegraphics[width=0.45\textwidth]{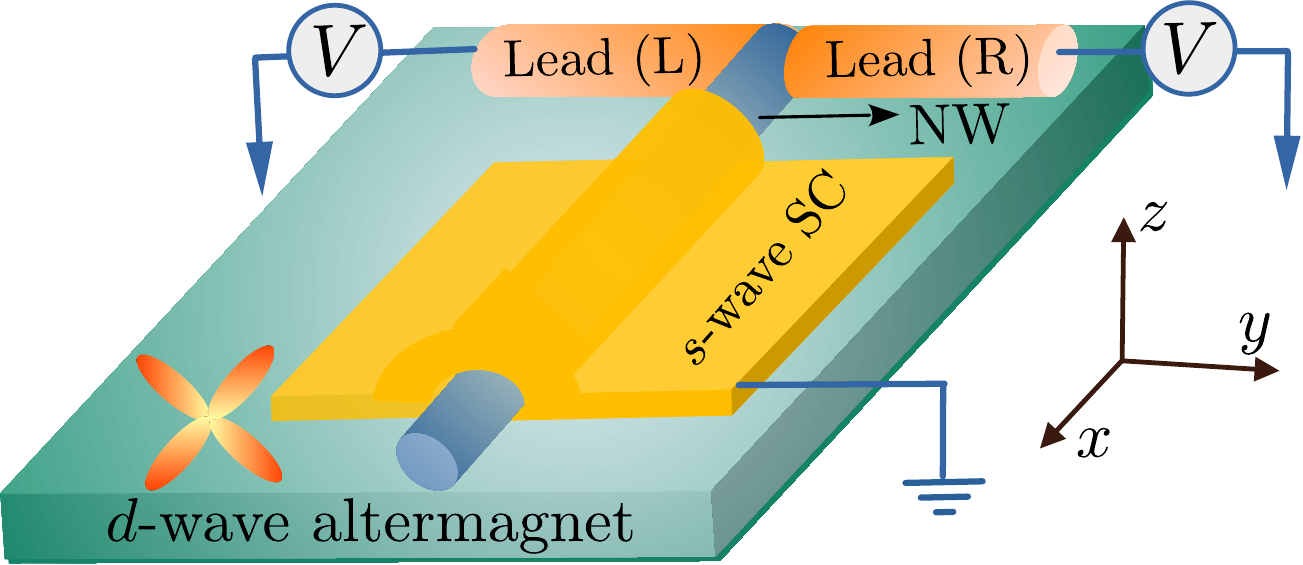}}
	\caption{Schematic diagram of our ``T''-junction setup for transport and shot noise signature study of MZMs and AZMs emerged in a heterostructure comprised of 1D Rashba NW~(blue), $d$-wave 
	altermagnet~(green), and $s$-wave superconductor~(yellow). Two arms of the ``T''-junction are made of metallic leads~(orange) via which a bias voltage $V$ is applied across the junction. 
	The NW constitutes the middle leg of the ``T''-junction and both altermagnetism and superconductivity have been induced in the NW via the proximity effect. 
	}
	\label{fig:1}
\end{figure}

\begin{figure}[]
	\centering
	\subfigure{\includegraphics[width=0.49\textwidth]{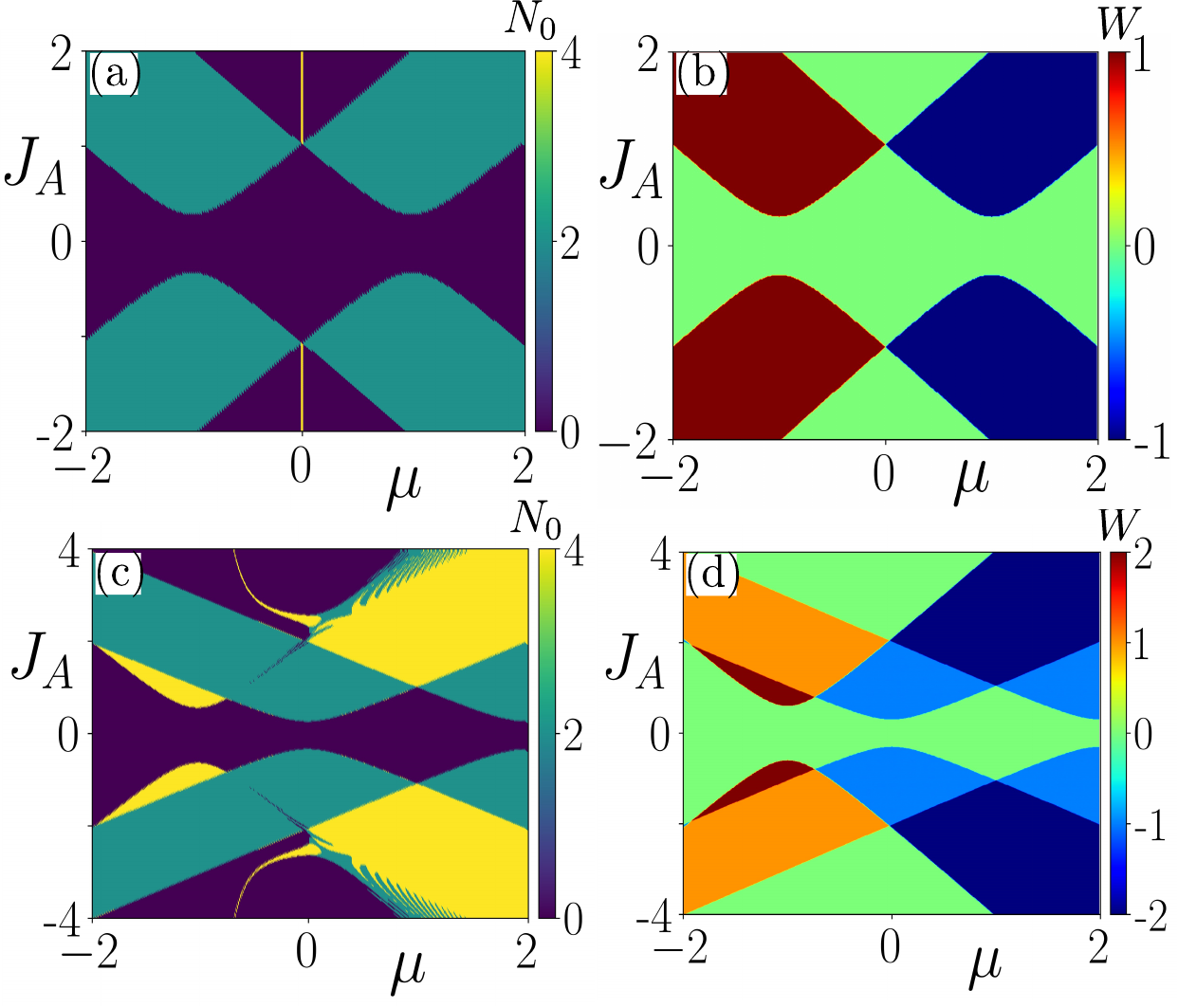}}
	\caption{In panel (a) we depict the number of zero energy modes $N_{0}$ obtained in OBC spectrum of Hamiltonian Eq.~(\ref{eq:Hamiltonian}) in $\mu$-$J_{A}$ plane choosing
	$\alpha=1,\beta=0$ and system size $N=300$ lattice sites. In panel (b) we portray the corresponding winding number $W$ computed using Eq.~(\ref{eq:winding_num}). We repeat (a) and (b) 
	in panels (c) and (d), respectively, for NNN model \ie $\alpha=\beta=1$, and $N=1000$ lattice sites. We consider other model parameters to be $(t_{h},\lambda_{R},\Delta)=~(1.0,0.5,0.3)$.}
	\label{fig:2}
\end{figure}

In this article, we begin with an alternative model based on 1D Rashba NW placed in closed proximity to a $s$-wave superconductor and a $d$-wave altermagnet~\cite{ghorashi2023altermagnetic}~
(see Fig.~\ref{fig:1} for schematic). Note that, proximity induced altermagnetism satisfies the criterion of broken TRS without suppressing the superconducting order due to net zero magnetization.
This setup inherits an effective 1D $p$-wave superconductivity hosting MZMs at the two ends of the NW akin to Kitaev model~\cite{ghorashi2023altermagnetic}. 
One can characterize these MZMs by calculating the appropriate topological index~(here winding number $W$). Interestingly, the concerned heterostructure also host four zero-energy modes with vanishing topological index~[see Fig.~\ref{fig:2}(a) and Fig.~\ref{fig:2}(b)]. We coin them as accidental zero modes~(AZMs). The latter is referred to as topological modes in earlier literature~\cite{ghorashi2023altermagnetic} without further investigation. Having extensively studied the transport on magnetic field/ferromagnet based TSC~\cite{HaimPRL2015,Haim_PRB2015}, we here examine the tunneling conductance and shot noise of this altermagnet based TSC setup. To be precise, we seek answers to the following intriguing questions: How to topologically distinguish the MZMs from 
the AZMs as far as their transport and shot noise signatures are concerned? What are the non-trivial signatures of multiple MZMs in the extended altermagnet TSC that does not exist in regular altermagnet TSC? 

\begin{figure}[]
	\centering
	\subfigure{\includegraphics[width=0.5\textwidth]{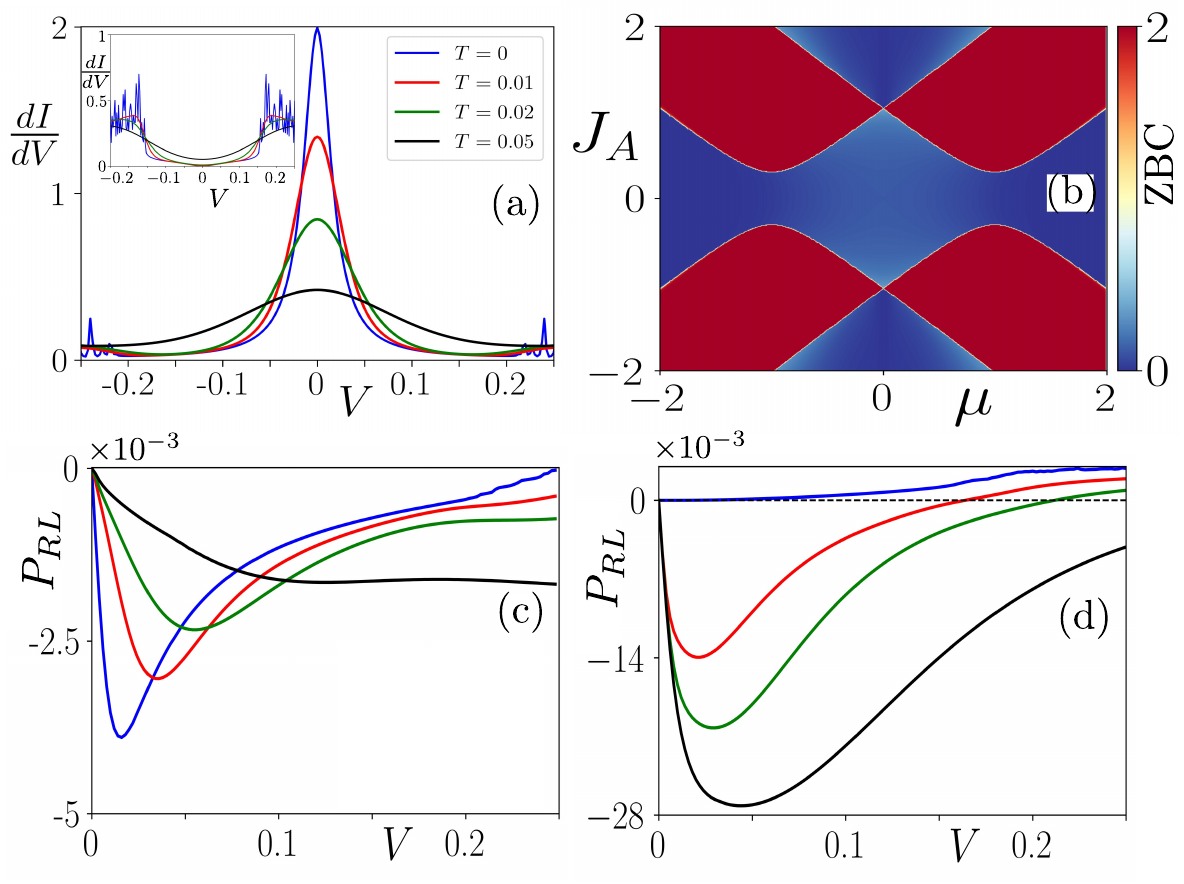}}
	\caption{Panel (a) represents differential conductance $\frac{dI}{dV}$ as a function of voltage bias $V$ choosing different temperatures for the nearest neighbor model~$\alpha=1,~\beta=0$ 
	and $|W|=1$ \ie MZMs.
	Inset of panel (a) stands for the same with $W=0$ \ie AZMs. At $T=0$, former exhibits a quantized peak of $2e^{2}/h$ at $V=0$ while latter does not exhibit any such peak. 
	In panel (b) we depict $\frac{dI}{dV}(V=0)$~(ZBC) in the plane of $\mu$ and $J_{A}$ at $T=0$. In panel (c) and (d) we illustrate the shot noise cross-correlation $P_{RL}$ corresponding to panel (a) 
	and its inset, respectively. 
	Here, we perform numerical computations for $N=300$ lattice sites. For panels (a) and (c) we choose $\mu=1.0,J_{A}=1.5$. On the other hand, for inset of panel (a) and panel (d) we consider 
	$\mu=0,J_{A}=1.2$. All the other model parameters remain same as mentioned in Fig.~\ref{fig:2}.
	}
	\label{fig:3}
\end{figure}

\begin{figure}[]
	\centering
	\subfigure{\includegraphics[width=0.5\textwidth]{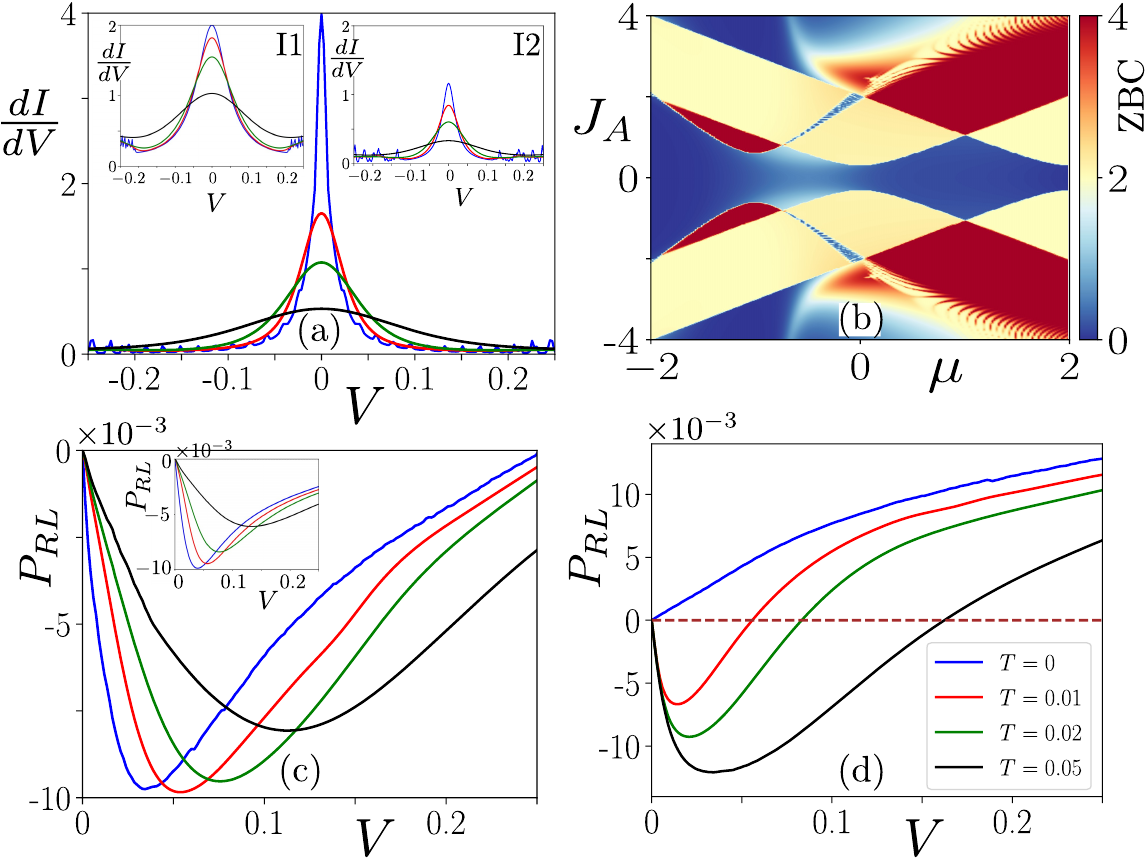}}
	\caption{
	In panel (a), we depict $\frac{dI}{dV}$ for MZMs with $|W|=2$ as a function of bias voltage choosing different temperatures. The two insets $I_{1}$, and $I_{2}$ showcase the same for MZMs with 
	$|W|=1$ and AZMs, respectively. In panel (b), we illustrate ZBC in $\mu$-$J_{A}$ plane at $T=0$. Clearly, at $T=0$ ZBC exhibits quantized value of $|W|\times 2 e^{2}/h$ for MZMs and non-
	quantized value for AZMs. Such quantization of ZBC disappears at $T\neq 0$ 
	for any kind of zero modes. Panel (c) and it's inset represents $P_{RL}$ for different temperatures corresponding to MZMs with $|W|=2$ and $|W|=1$, respectively. 
	On the other hand, in panel (d) we show $P_{RL}$ corresponding to AZMs. 
	We perform our numerical computation considering $N=1000$ lattice sites. For panel (a) we choose $\mu=1.0$, $J_{A}=1.2$. For the inset I1~(I2) we choose $\mu=0.0$, $J_{A}=1.0$~($\mu=-0.5$, 
	$J_{A}=3.0$). All the other model parameters remain same as mentioned in Fig.~\ref{fig:2}.
	}
	\label{fig:4}
\end{figure}

To answer the first question, we consider a three terminal ``T''-shaped junction~(see Fig.~\ref{fig:1} for schematic). We obtain zero bias conductance~(ZBC), $\frac{dI}{dV} (V=0)$ to be quantized with value $2e^{2}/h$ for MZMs and some non-universal value for AZMs~[see Fig.~\ref{fig:3}(a) and the inset of Fig.~\ref{fig:3}(a), respectively]. However, at finite temperatures ZBC loses its quantization even for MZMs. On the other hand, for MZMs, shot noise becomes negative for all $V$ (less than the bulk superconducting gap) and approaches to zero as power law $-1/V$. This behavior is universal and remains valid up to a certain finite temperature. In sharp contrast, for AZMs, this behavior is non-universal [see Fig.~\ref{fig:3}(c) and Fig.~\ref{fig:3}(d), respectively]. To address the second question, 
we incorporate a second nearest neighbor term in both the hopping and SOC in our model to obtain multiple MZMs. However, in this case too we cannot get rid of the AZMs with null topological index. For MZMs with winding number $W$, we obtain $|W|\times 2 e^{2}/h$ quantized ZBC and no such quantization appears for AZMs at zero-temperature~[see Fig.~\ref{fig:4}(a) and Fig.~\ref{fig:4}(b)]. Corresponding shot noise signatures are also similar like the previous case~[see Fig.~\ref{fig:4}(c) and Fig.~\ref{fig:4}(d)], thus distinguishing between the topological MZMs and non-topological AZMs. 

\textcolor{blue}{\textit{Model.}---} We begin with a 1D semiconducting nanowire~(NW) with Rashba SOC of relativistic origin. The NW is placed in close proximity to a $d$-wave altermagnet and an $s$-wave superconductor~(see Fig.~\ref{fig:1} for schematic). Superconductivity and altermagnetism are assumed to be induced into the NW as a consequence of proximity effect. We consider the following Bogoliubov-de Gennes~(BdG) basis: $\Psi_{k}=\left\{\psi_{k\ua},\psi_{k\da},\psi_{-k\da}^{\dagger},-\psi_{-k\ua}^{\dagger} \right\}^{\textbf{t}}$ with $\psi_{k\ua}^{\dagger}$~($\psi_{k\ua}$) and $\psi_{k\da}^{\dagger}$~($\psi_{k\da}$) represents the creation~(annihilation) operator for spin up~(down) sector respectively, and $\mathbf{t}$ stands for transpose operation. Thus the Hamiltonian is given by 
$H=\sum_{k} \Psi_{k}^{\dagger}H(k)\Psi_{k}$ with
\begin{eqnarray}
H(k)=&&\left[t_{h}\left(\alpha~\cos(k)+ \beta ~\cos(2k)\right)-\mu\right]\tau_{z}\sigma_{0}\non\\
&&+\left[\lambda_{R} \left(\alpha~\sin(k)+ \beta ~\sin(2k)\right)\right]\tau_{z}\sigma_{y}\non\\
&&+J_{A}\cos(k)\tau_{0}\sigma_{z}+\Delta \tau_{x}\sigma_{0} \label{eq:Hamiltonian}\ ,
\end{eqnarray}
where $t_{h}$, $\mu$, $\lambda_{R}$, $J_{A}$, and $\Delta$ denote the hopping amplitude, chemical potential, Rashba SOC strength, proximity induced altermagnetic strength and induced superconducting gap, respectively. The Pauli matrix $\vect{\tau}$~($\vect{\sigma}$) stands for the particle-hole~(spin) degrees of freedom. Note that, $\alpha=1$ and $\beta$ is assumed to be either 0 or 1. Former case represents the model with only nearest neighbor hopping and SOC~\cite{ghorashi2023altermagnetic} while latter stands for the next nearest neighbor (NNN) one. We  comprehensively study the bulk band structure of this Hamiltonian in the supplementary material (SM)~\cite{supp} and discuss the band structure of various topological phases (gapped phase) as well as the phase boundaries (gapless phase). The Hamiltonian~[Eq.~(\ref{eq:Hamiltonian})] preserves chiral symmetry $S=\tau_{y} \sigma_{y}$: $S^{-1}H(k)S=-H(k)$, and particle-hole symmetry $\cal{C} $$=  \tau_{y} \sigma_{y} \cal{K} $: ${\cal{C}}^{-1}H(k)\cal{C} = $$-H(-k)$. Hence, the system possesses an ``effective" or pseudo TRS, given by ${\cal{T}}^{\prime}= {\cal{K}}$:  ${\cal{T}}^{\prime -1} H(k){\cal{T}}^{\prime} =H(-k)$ and belongs to BDI topological class~\cite{TewariPRL2012,ghorashi2023altermagnetic}. Note that, the proximity induced altermagnetic term in the Hamiltonian breaks the actual TRS $\cal{T}$$=i\tau_{0} \sigma_{y} \cal{K}$: ${\cal{T}}^{-1}H(k)\cal{T} \neq$ $ H(-k)$.

Exploiting chiral symmetry of the system, one can topologically characterize this system by computing the $Z$ index or winding number $W$ in the following way. As chiral symmetry operator $S$ anti-commutes 
with the Hamiltonian $H(k)$, the chiral basis $U_{s}$ diagonalizing $S$ makes the $H(k)$ in anti-diagonal block form as:
\begin{eqnarray}
U_{s}^{\dagger}H(k)U_{s}=\begin{pmatrix}
0&q(k)\\
q^{\dagger}(k) &0\\
\end{pmatrix} \label{eq:anti_block},
\end{eqnarray}
with $2\times2$ block $q(k)$ is given by
\begin{widetext}
\begin{eqnarray}
q(k)=\!\!\begin{pmatrix}
i \lambda_{R}\left[\alpha~\sin(k)+ \beta ~\sin(2k)\right]-\Delta ~&~t_{h}\left[\alpha~\cos(k)+ \beta ~\cos(2k)\right]-\mu+ J_{A}\cos(k)\\
t_{h}\left[\alpha~\cos(k)+ \beta ~\cos(2k)\right]-\mu-J_{A}\cos(k)& -i \lambda_{R}\left[\alpha~\sin(k)+ \beta ~\sin(2k)\right]+\Delta
\end{pmatrix}. \label{eq:qk}
\end{eqnarray} 
\end{widetext}
The corresponding winding number is given by~\cite{RyuNJP2010,ChiuRMP2016,Mondal2023_NW}
\begin{eqnarray}
W=\frac{i}{2\pi} \int_{\rm{BZ}} dk \operatorname{Tr}\left[q^{-1}(k) \partial_{k} q(k)\right] \label{eq:winding_num}.
\end{eqnarray}

\textcolor{blue}{\textit{Zero-energy modes and their topological characterization}---}First, we consider the nearest neighbor model~\cite{ghorashi2023altermagnetic} with $\alpha=1,~\beta=0$ in Hamiltonian~[Eq.~(\ref{eq:Hamiltonian})]. We diagonalize the corresponding Hamiltonian choosing open boundary condition~(OBC) to obtain the real space energy eigenvalues. From there we count the number of zero energy end modes $N_{0}$ and depict them in $\mu$-$J_{A}$ plane in Fig.~\ref{fig:2}(a). We obtain $N_{0}=2$ within the regions bounded by the bulk gap closing lines: $J_{A}^{2}=\left(t_{h}-\mu \right)^{2} +\Delta^{2}$ and $J_{A}^{2}=\left(t_{h}+\mu \right)^{2} +\Delta^{2}$~[see SM~\cite{supp} for details of gap structure]. Interestingly, $N_{0}$ becomes $4$ on the lines given by 
$\mu=0$ for $|J_{A}/t_{h}|>1$. These in-gaped zero-energy modes emerge without any bulk gap closing as far as changing $\mu$ is concerned. To be precise, for any $|J_{A}/t_{h}|>1$, if one modulates $\mu$ from small negative to positive value, these modes suddenly arise at $\mu=0$ without any bulk gap closing.

Using Eq.~(\ref{eq:winding_num}) we compute the corresponding winding number $W$ and depict in $\mu$-$J_{A}$ plane in Fig.~\ref{fig:2}(b). We obtain non-trivial value of winding number~($W=\pm 1$) for the zero energy modes bounded by bulk gap closing lines $J_{A}^{2}=\left(t_{h}-\mu \right)^{2} +\Delta^{2}$ and $J_{A}^{2}=\left(t_{h}+\mu \right)^{2} +\Delta^{2}$. It clearly confirms them to be topological \ie MZMs. However, vanishing $W$ for other type of zero modes tags them as topologically trivial. 
We coin them as AZMs as mentioned before.

\textcolor{blue}{\textit{Transport and shot noise signature}---} We note that both the MZMs and AZMs are localized at the end of the NW~(see SM \cite{supp} for localization properties). To distinguish between them, we adopt transport and shot noise study following the Landauer-B{\"u}ttiker formalism~\cite{BLANTER20001,Anantram1996}. In order to do that, we consider a three terminal ``T''-shaped junction~(see Fig.~\ref{fig:1})~\cite{HaimPRL2015,Haim_PRB2015} with two arms connected to two metallic leads~(left lead, $L$ and right lead, $R$), maintained at bias voltage $V$ and the middle leg~(the proximity induced NW) is grounded. Interestingly, in this ``T''-shaped junction, the current-current cross correlation~($P_{RL}$) between the leads $L$ and $R$ is negative and approaches to bulk gap as $-1/V$ for MZMs. This behavior is universal, even valid at finite temperatures, and robust against disorder~\cite{Haim_PRB2015}. However, this universal feature is not applicable for topologically trivial zero modes. For this reason, $P_{RL}$ serves as an appropriate quantity for characterizing MZMs in transport study, in addition to differential conductance. If $M$ number of sites of the NW is connected to the leads, each of the leads individually contains $2M$ number of transverse channels encompassing both the spin components~($\ua, \da$)~\cite{Haim_PRB2015}. For energies below proximity induced superconducting gap~($E<\Delta$), scattering process occurs only via reflection~[both normal and Andreev refection (AR)] and corresponding scattering matrix is given by the reflection matrix~\cite{Beenakker1991,Haim_PRB2015,Fisher1981,IIDA1990219}
\begin{eqnarray}
	 r_{\rm{tot}}(E)&=&\begin{pmatrix}
	 r^{ee}(E)&r^{eh}(E)\\
	 r^{he}(E)&r^{hh}(E)
	 \end{pmatrix} \non\\
  &=&1-2 \pi i W_{c}^{\dagger} \left[E - H_{\rm{TB}}+i \pi W_{c} W_{c}^{\dagger} \right] W_{c}\ ,\label{eq:rtot_main}
\end{eqnarray} 
where $r^{\alpha \beta}$ is the refection matrix block for $\beta$ type incoming particles to be reflected as $\alpha$ with $\{\alpha,\beta\} \in \{e,h\}$, and $H_{\rm{TB}}$ is the tight binding Hamiltonian of the NW in OBC, and $W_{c}$ corresponds to a energy independent matrix in wide band limit of the metallic leads and carrying the information of coupling between NW and leads~(see SM \cite{supp} for the details). Note that, each block of $r_{\rm{tot}}$ is a $4M\times4M$ matrix. In terms of blocks of $r_{\rm{tot}}$ matrix, total current $I$ through the leads ($L$ and $R$) and their corresponding cross correlation 
$P_{RL}$ can be written as~\cite{Anantram1996,Haim_PRB2015}~(see~\cite{supp} for the derivation)
\begin{eqnarray}
I&=&\frac{2e}{h}\int_{0}^{\infty} dE \,\, \operatorname{Tr}\left[r^{he} r^{he \dagger}\right]  f_{e}(E) \label{eq:current_finite_main}\ ,\\
P_{RL}&=&\frac{e^{2}}{h}\!\! \sum_{i\in R, j\in L} \int_{0}^\infty dE \sum_{\G,\D} \Pc^{\G\D}_{ij}(E) f_{\G}(E)(1-f_\D(E))\ ,  \non \\ \label{eq:noise_finite_main}
\end{eqnarray}
where,\,$\Pc^{\G\D}_{ij}(E) \!\!=\!\! \sum_{k,l}\! \sum_{\A,\B} \sgn(\A)\sgn(\B) A^{\G\D}_{kl}(i,\!\A) A^{\D\G}_{lk}(j,\B)$
and \hskip 0.2cm  $A^{\G\D}_{kl}(i,\A,E) = \delta_{ik} \delta_{il} \delta_{\A\G} \delta_{\A\D} - (r^{\A\G}_{ik})^* r^{\A\D}_{il}$. Here, $k,l=1,...,4M$ and run over all the transverse channels in left and right lead and $\A,\B\in\{e,h\}$.  Explicit expressions of $\Pc^{\G\D}_{i,j}(E)$ in terms of the reflection matrix is presented in~\cite{supp}.
 Also, $f_{e}(E)=1- f_{h}(-E)=1/\left\{ \exp{(E-eV)/k_{B}T}\right\}$ is the Fermi-Dirac distribution function of incoming electrons in the leads and $T$ is the temperature ($k_B=1$). At zero temperature the above two equations read~\cite{Haim_PRB2015,Nilsson2008} $I=\frac{2e}{h}\int_{0}^{eV} dE~ \operatorname{Tr}\left[r^{he}(E)~ r^{he \dagger}(E)\right]$ and $P_{RL}=\frac{e^{2}}{h}\sum_{i\in R,j\in L} \int_{0}^{eV} dE \hspace*{2 mm} {\cal{P}}_{ij}^{eh}(E)$.

We compute the differential conductance $\frac{dI}{dV}$ for MZMs~(AZMs) and depict them as a function of the bias voltage $V$ in Fig.~\ref{fig:3}(a)~[inset of Fig.~\ref{fig:3}(a)] for different temperature $T$ scales. At $T=0$, we obtain quantized value of $2e^{2}/h$ as ZBC signature for MZMs. Although no such quantization appears for AZMs. Note that, the small peaks at high bias voltage in $\frac{dI}{dV}$ in Fig~\ref{fig:3}(a) and its inset originate from the bulk states above the gap. We also illustrate ZBC in $\mu$-$J_{A}$ plane in Fig.~\ref{fig:3}(b) to have better understanding 
of the transport signature at $T=0$. Note that, $\frac{dI}{dV}$ follows the behavior of $W$ as depicted in Fig.~\ref{fig:2}(b). Also, AZMs exhibit vanishingly small $\frac{dI}{dV}$ at $\mu=0$ lines for 
$|J_{A}/t_{h}|>1$. However, at $T\neq 0$, quantization is lost even for MZMs. Hence, for finite temperature $\frac{dI}{dV}$ doesn't play the feasible role to distinguish between MZMs and AZMs. In this scenario shot noise plays a pivotal role. We compute shot noise $P_{RL}$ for both the types of zero modes. For MZMs, within the bulk gap, we obtain $P_{RL}$ to be negative and tends to zero as $-1/V$ nature~[see Fig.~\ref{fig:3}(c)] while for AZMs $P_{RL}$ is positive for $T=0$ and crossing over from negative to positive value~[see Fig.~\ref{fig:3}(d)]. Hence, transport and shot noise study clearly distinguish between MZMs and AZMs appeared in our concerned heterostructure with nearest neighbor connection.

\textcolor{blue}{\textit{Next nearest neighbor model}---}
Although our results for nearest neighbor model are conclusive, we extend our investigation to examine the applicability of the above study in presence of multiple MZMs. To accomplish this, we consider NNN model with $\alpha=\beta=1$ in Eq.~(\ref{eq:Hamiltonian}). At first, we compute eigenvalue spectra obtained employing OBC. From there we count the number of zero energy modes $N_{0}$ and illustrate it in $\mu$-$J_{A}$ plane in Fig.~\ref{fig:2}(c). We obtain $N_{0}=2$ and $N_{0}=4$ within the regions bounded by gap closing lines: $J_{A}^{2}=\left(2~t_{h}-\mu \right)^{2} +\Delta^{2}$, $J_{A}^{2}=\mu^{2} +\Delta^{2}$, and $J_{A}^{2}=4\left[\left(t_{h}+\mu \right)^{2} +\Delta^{2}\right]$~(see SM~\cite{supp} for details of the gap structure). Therefore, in this extended model, we as well encounter zero modes with $N_{0}=4$ having no such gap closing origin~[see horn-shaped yellow region in Fig.~\ref{fig:2}(c))]. We compute winding number using Eq.~(\ref{eq:winding_num}) and demonstrate them in $\mu$-$J_{A}$ plane in Fig.~\ref{fig:2}(d). We obtain $W=\pm 1$~($W=\pm 2$) for zero modes with $N_{0}=2$~($N_{0}=4$) associated with bulk gap closing origin. Here also the zero modes without gap closing origin exhibit null winding number. Hence, it is  evident that 
the latter types of zero modes coined as AZMs are non-topological in nature.

We repeat our transport and shot noise study for this model and illustrate them in Fig.~\ref{fig:4}. In this scenario, we compute $\frac{dI}{dV}$ for three cases: MZMs with $|W|=2$, $|W|=1$, and AZMs with $W=0$. 
The corresponding behavior is displayed in Fig.~\ref{fig:4}(a) and its insets $I_{1}$ and $I_{2}$, respectively. We also illustrate ZBC in $\mu$-$J_{A}$ plane in Fig.~\ref{fig:4}(b). The prong-like structure in Fig.~\ref{fig:4}(b) exhibits the mixture of $2 e^{2}/h$ and $4 e^{2}/h$ quantization as a consequence of finite size effect~(see SM~\cite{supp} for the details). At $T=0$, we obtain 
ZBC to be quantized with the peak height of $4 e^{2}/h$~($2e^{2}/h$) for MZMs with $|W|=2$~($|W|=1$). On the other hand, non-universal ZBC appears for AZMs at $T=0$ irrespective of their number. 
At finite temperature, ZBC lose its quantization for MZMs alike the previous case~[see Fig.~\ref{fig:4}(a) and its insets $I_{1}$ and $I_{2}$]. 
Hence, at $T=0$ in this model too, ZBC provides conclusive evidence to distinguish between the multiple MZMs and AZMs. However, for $T \neq 0$ we again rely on the behavior of current cross-correlation $P_{RL}$. We compute $P_{RL}$ for the three cases mentioned before and depict them as a function of $V$ in Fig.~\ref{fig:4}(c), inset of Fig.~\ref{fig:4}(c) and Fig.~\ref{fig:4}(d), respectively. Akin to the nearest-neighbor case, here also we obtain $P_{RL}$ to be negative~(negative to positive transition) for MZMs~(AZMs) within the bulk TSC gap. Hence, for this extended model as well, 
our transport and shot noise study can clearly distinguish between the topological MZMs and non-topological AZMs.

\textcolor{blue}{\textit{Summary and conclusion}---} 
To summarize, in this article, we consider a heterostructure comprised of a 1D Rashba NW in close proximity to a bulk $s$-wave superconductor and a TRS broken $d$-wave altermagnet. This setup anchors topological MZMs in addition to non-topological AZMs, localized at the end of the NW. To differentiate between MZMs and AZMs, we perform transport and shot noise study in a three terminal setup that is capable of realizing current cross correlations. We repeat our investigations in a NNN hopping model to benchmark our findings even in the presence of multiple MZMs. We investigate the stability of transport results against static random onsite disorder~(see~\cite{supp} for details). Importantly, the AZMs are referred as topological MZMs in the Ref.~\cite{ghorashi2023altermagnetic} without providing appropriate topological characterization. However, in contrast, by computing winding number, and performing transport-shot noise study, we argue that these AZMs are non-topological in nature both in nearest-neighbor and NNN model.

Due to resonant Andreev reflection, a single MZM exhibits $2e^{2}/h$ quantized ZBC at zero temperature in a normal metal-superconductor junction~\cite{PLeePRL2009}. Nevertheless, experimental detection of such quantized response as indirect signature of MZMs is still debatable as such ZBC peak can also have other physical origins such as Kondo effect~\cite{GoldhaberNature1998, KondoScience1998}, disorder~\cite{AlexanderPRL2012}, and Andreev bound state~\cite{BrouwerABS2012,Lee2014,ABS_PRB2019} etc. It is also challenging as it requires temperature to be sufficiently smaller than the width of the ZBC peak of MZMs. At higher temperatures, the height of the ZBC peak reduces to a non-quantized value as a result of thermal broadening. 
Interestingly, this ambiguity can be resolved by probing current cross correlation~($P_{RL}$) in three terminal junction setup~\cite{HaimPRL2015,Haim_PRB2015,Smirnov2022}. For MZMs, $P_{RL}$ is negative closed to zero bias voltage as a consequence of perfect Andreev reflection, and approaches to zero as $-1/V$~(inside the bulk gap) stemming from non-local nature of MZMs, localized at the two ends of NW~\cite{Haim_PRL_2015, Haim_PRB2015}. This behavior is universal and persists even at finite temperatures~\cite{Haim_PRB2015}. Note that, there is an interplay among bulk gap, 
$V$, and $T$: as $T$ increases, the $-1/V$ behavior starts at larger $V$, but $V$ must still lie within the bulk TSC gap. In Fig.~\ref{fig:3}(c), for $T = 0.05$, the starting point of the $-1/V$ behavior is above the bulk gap, where the $-1/V$ power law no longer holds, leading to a deviation. However, for any non-topological zero modes, this feature is non-universal. Hence, we utilize these signatures in our altermagnetic setup to distinguish between MZMs and AZMs. We have explicitly checked that our observation remains unaltered even in the presence of multiple MZMs. The  distinction between topological MZMs and non-topological modes, such as Andreev bound states, has been previously studied using transport and shot noise signatures in the Rashba NW model~\cite{HaimPRL2015,Haim_PRB2015}. This clearly demonstrates the $-1/V$ dependence of shot noise due to MZMs. We have obtained similar behavior for shot noise in our altermagnet based hybrid setup.

We also construct a two-terminal junction with a TSC nanowire and attach two leads at the ends of it to compute the ZBC and check for its quantization. We utilize the Blonder-Tinkham-Klapwijk (BTK) formalism~\cite{BTK_NS1982} to compute the differential conductance and perform numerical simulations using the Python package KWANT~\cite{KWANT_Groth_2014} to 
determine the differential conductance. Our results exhibit excellent agreement with the ``T''-junction results (see~\cite{supp} for detailed analysis).


Throughout our study, we set $\beta = \alpha$. However, we show analytically that any $\beta$ value in the range $|\alpha/2| \leq |\beta| \leq |\alpha|$ yields the same qualitative results, as verified numerically for $\beta = 0.6 \alpha$ (see SM~\cite{supp}). Similarly, while $\lambda_{R} = 0.5 t_h$ corresponds to a strong SOC value, we demonstrate that any non-zero $\lambda_{R}$ gives rise to identical results (see SM~\cite{supp}). Very recently, Liu et. al. discuss the absence of spin-splitting in ${\rm{RuO_{2}}}$~\cite{Liu_2024_PRL}. Hence, MnTe~\cite{Amin2024,Krempaský2024} 
can be the preferable candidate as a $d$-wave altermagnet in our setup.

Note that, topological MZMs can also be characterized with the help of Majorana polarization~\cite{Sticlet2012,Sedlmayr_2015,SedlmaryrPRB2015,BENA2017349,Glodzik2020,awoga2024}, Majorana entropy~\cite{Smirnov2015,Smirnov2021}, and several Josephson junction based signatures~\cite{Cayao2015,Awoga2019,Cayao2021,Baldo_2023} in our setup. In future, one may apply external periodic drive to this setup to investigate the emergence and behavior of AZMs. Our study confirms the existence of MZMs and AZMs in long range version of the concerned model as well. Hence, the route to engineer MZMs without the appearance of AZMs in an altermagnetic heterostructure can be a future direction as well.

\textcolor{blue}{\textit{Acknowledgements}---}D.M., A.P., and A.S. acknowledge SAMKHYA: High-Performance Computing Facility provided by Institute of Physics, Bhubaneswar and the Workstation provided by Institute of Physics, Bhubaneswar from DAE APEX Project, for numerical computations. D.M. thanks Pritam Chatterjee for stimulating discussions. T.N. acknowledges NFSG from
BITS Pilani-NFSG/HYD/2023/H0911.

\bibliography{bibfile}{}


\clearpage
\newpage

\begin{onecolumngrid}
	\begin{center}
		{\fontsize{12}{12}\selectfont
			\textbf{Supplementary Material for ``Distinguishing between topological and trivial zero modes via transport and shot noise study in an altermagnet heterostructure''\\[5mm]}}
		{\normalsize Debashish Mondal\orcidD{},$^{1,2}$ Amartya Pal\orcidA,$^{1,2}$  Arijit Saha\orcidC{},$^{1,2}$ and Tanay Nag\orcidB{},$^{3}$\\[1mm]}
		{\small $^1$\textit{Institute of Physics, Sachivalaya Marg, Bhubaneswar-751005, India}\\[0.5mm]}
		{\small $^2$\textit{Homi Bhabha National Institute, Training School Complex, Anushakti Nagar, Mumbai 400094, India}\\[0.5mm]}
		{\small $^3$\textit{Department of Physics, BITS Pilani-Hydrabad Campus, Telangana 500078, India}\\[0.5mm]}
		{}
	\end{center}
	
	\newcounter{defcounter}
	\setcounter{defcounter}{0}
	\setcounter{equation}{0}
	\renewcommand{\theequation}{S\arabic{equation}}
	\setcounter{figure}{0}
	\renewcommand{\thefigure}{S\arabic{figure}}
	\setcounter{page}{1}
	\pagenumbering{roman}
	
	\renewcommand{\thesection}{S\arabic{section}}
	
         \tableofcontents 


\pagebreak
\section{End localization of zero energy modes} \label{sec:LDOS}
In the main text, we mention that, both type of zero energy modes~[Majorana zero modes~(MZMs) and Accidental zero modes~(AZMs)~] are localized at the end of the nanowire~(NW). To establish that statement, we compute local density of states~(LDOS) corresponding to zero energy modes emerged in our setup and depict them as a function of site index $x$ in Fig.~\ref{fig:S1}. For all the cases, zero energy modes are localized at the ends of the NW. Note that, LDOS is normalized with its maximum value.
 \begin{figure}[H]
	\centering
	\subfigure{\includegraphics[width=0.9\textwidth]{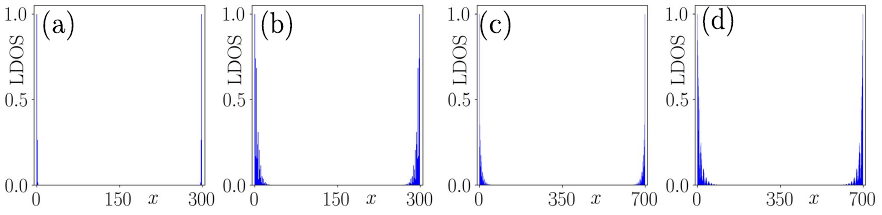}}
	\caption{(a)~[(b)] Local density of states~(LDOS) profile for AZMs~[MZMs with $|W|=1$] emerged in the nearest neighbor model~($\alpha=1,~\beta=0$) is shown as a function of site index. 
	In panel (c)~[(d)] LDOS distribution is depicted as a function of site index for AZMs~[MZMs with $|W|=2$] appeared in the next nearest neighbor model~($\alpha=\beta=1$). We choose the model 
	parameters $(\mu,J_{A},N)$ to be (0.0,1.2,300), (1.0,1.5,300), (-0.5,3.0,700), and (1.0,1.2,700) for panels (a), (b), (c), and (d), respectively. All other model parameters remain same as mentioned in 
	Fig.~2 of the main text.
	}
	\label{fig:S1}
\end{figure}

\section{Bulk gap study} \label{sec:bulk}
\begin{figure}[h]
	\centering
	\subfigure{\includegraphics[width=0.8\textwidth]{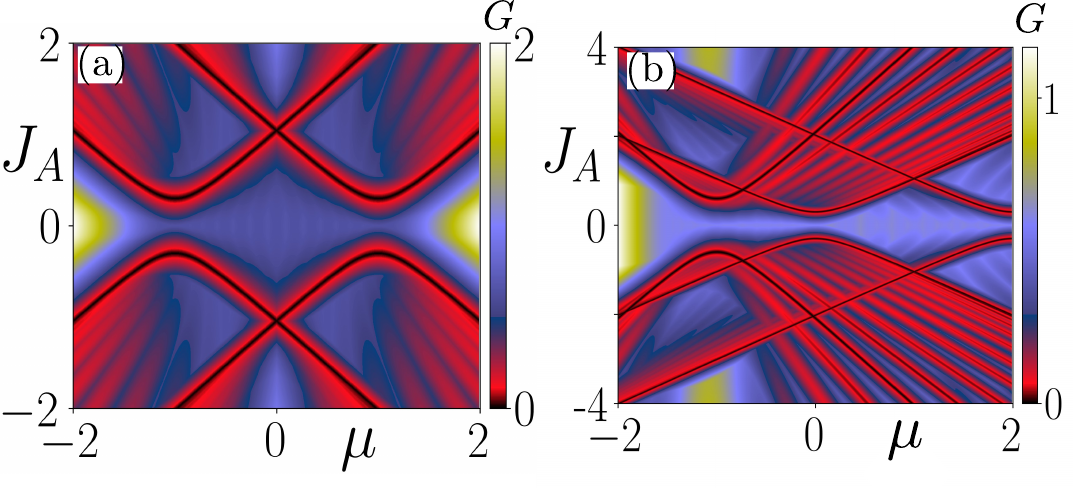}}
	\caption{(a) Bulk gap $G$ for nearest neighbor model~($\alpha=1,~\beta=0$). (b) Bulk gap $G$ for nearest neighbor model~($\alpha=\beta=1$). Black lines represent gapless lines i.e $G=0$. All the model parameters remain same as Fig.~2 in the main text.
	}
	\label{fig:S2}
\end{figure}
In the main text, we compute the number of zero energy modes and winding number to topologically characterize the system. Based on that we conclude the existence of AZMs with vanishing winding number. In general, a topological phase transition is associated with the bulk gap closing in the band structure of the system. In this section, we show that AZMs appear without bulk gap closing which further strengthen our statement regarding the non-topological character of AZMs. To this end, we numerically compute the bulk gap $G$ of the eigenvalue spectrum of the Hamiltonian~[Eq.~(1) in main text] for nearest neighbor (NN)~($\alpha=1,~\beta=0$) and next nearest neighbor (NNN)~($\alpha=\beta=1$) model. We depict them in $\mu$-$J_{A}$ plane in Fig.~\ref{fig:S2}(a), and Fig.~\ref{fig:S2}(b), respectively. For both the panels, the black lines represent gapless lines~($G=0$). Note that, the red regions are gapped even with vanishingly small bulk gap. Interestingly, for NN model, with any constant line $|J_A|>1$, there is no bulk gap closing near $\mu =0$ [Fig.\,\ref{fig:S2}(a)]. However, one can observe the appearance of zero energy modes in the spectrum [see Fig.\,2(a) of the main text]. Therefore, we can label them as trivial AZMs. Similar observation can be made comparing Fig.\,\ref{fig:S2}(b) and Fig.\,2(c) (see the main text).
\subsection{Analytical expressions for gapless lines} \label{subsec:ana_gap}
To have analytical expressions for the gapless lines, we utilize the chiral symmetry of the system. Diagonal basis $U_{s}$ for chiral operator transforms system Hamiltonian into an anti-diagonal 
block form:

\begin{eqnarray}
U_{s}^{\dagger}H(k)U_{s}=\begin{pmatrix}
0&q(k)\\
q^{\dagger}(k) &0\\
\end{pmatrix} \label{eq:anti_block},
\end{eqnarray}
with $2\times2$ block $q(k)$ is given by

\begin{eqnarray}
q(k)=\!\!\begin{pmatrix}
i \lambda_{R}\left[\alpha~\sin(k)+ \beta ~\sin(2k)\right]-\Delta ~&~t_{h}\left[\alpha~\cos(k)+ \beta ~\cos(2k)\right]-\mu+ J_{A}\cos(k)\\
t_{h}\left[\alpha~\cos(k)+ \beta ~\cos(2k)\right]-\mu-J_{A}\cos(k)& -i \lambda_{R}\left[\alpha~\sin(k)+ \beta ~\sin(2k)\right]+\Delta
\end{pmatrix} \label{eq:qk_supp}.
\end{eqnarray}
If gap closing occurs for $k=k_{*}$, then $\rm{det}[q(k_{*})]=0$. $\forall~ \alpha,~\beta \in \mathbb{R}$, the real solutions of $k_{*}$ are given by
\begin{eqnarray}
&&\alpha~\sin(k_{*})+ \beta ~\sin(2k_{*})=0\ , \non\\
 &\implies &\sin(k_{*})~\left[ \alpha+2\beta~ \cos(k_{*})\right]=0\ . \label{eq:sup_condition}
\end{eqnarray}

Here, the first term $\sin(k_{*})=0$ yields $k_{*}=0~\text{or}~ \pi$. Now, $\rm{det}[q(0)]=0$ gives $J_{A}^{2}=\left[(\alpha+\beta)t_{h}-\mu\right]^{2}+\Delta^{2}$ while $\det[q(\pi)]=0$ provides $J_{A}^{2}=\left[(\alpha-\beta)t_{h}+\mu\right]^{2}+\Delta^{2}$.
Hence, for the second term we can have three situations: 1. $\alpha = 0,\beta \neq 0$. 
Then there exists a solution given by $\cos(k_{*})=0\implies k_{*}=\pi/2 \implies \cos(2k_{*})=-1$ yielding gap-closing line $J_{A}^{2}=\left[\beta t_{h}+\mu\right]^{2}+\Delta^{2}$. 2. $\alpha \neq 0, \beta=0$. Then the second term has no solution. 3. $\forall~ \alpha, \beta \neq 0$, $\alpha + 2 \beta \cos(k_{*})=0  \implies \cos(k_{*})=-\frac{\alpha}{2\beta}$ and $\cos(2k_{*})=2 \cos^{2}(k_{*})-1=2(\frac{\alpha}{2\beta})^{2}-1$. Then gap closing condition is given by

\begin{eqnarray}
J_{A}^{2}\left(\frac{\alpha}{2\beta}\right)^{2}&=&\left[\alpha\left(-\frac{\alpha}{2\beta}\right)t_{h}+\beta\left\{2\left(-\frac{\alpha}{2\beta}\right)^{2}-1 \right\} t_{h}-\mu\right]^{2}+\Delta^{2} \non\\
\implies J_{A}^{2}\left(\frac{\alpha}{2\beta}\right)^{2}&=&\left[-\beta t_{h}-\mu \right]^{2}+\Delta^{2} \non \\
\implies J_{A}^{2}&=&\left(\frac{2\beta}{\alpha}\right)^{2}\left[\left(\beta t_{h}+\mu \right)^{2}+\Delta^{2}\right]~~\forall~\alpha,\beta \neq 0\ . \label{eq:sup_gapless2}
\end{eqnarray}

Hence, for NN model gap closing lines are given by: $J_{A}^{2}=\left(t_{h}-\mu \right)^{2} +\Delta^{2}$ and $J_{A}^{2}=\left(t_{h}+\mu \right)^{2} +\Delta^{2}$. Also, for the NNN model: $J_{A}^{2}=\left(2~t_{h}-\mu \right)^{2} +\Delta^{2}$, $J_{A}^{2}=\mu^{2} +\Delta^{2}$, and $J_{A}^{2}=4\left[\left(t_{h}+\mu \right)^{2} +\Delta^{2}\right]$.


\section{Study of Bulk Band structure}\label{sec:band_structure}
\begin{figure}
	\centering
	\subfigure{\includegraphics[width=0.8\textwidth]{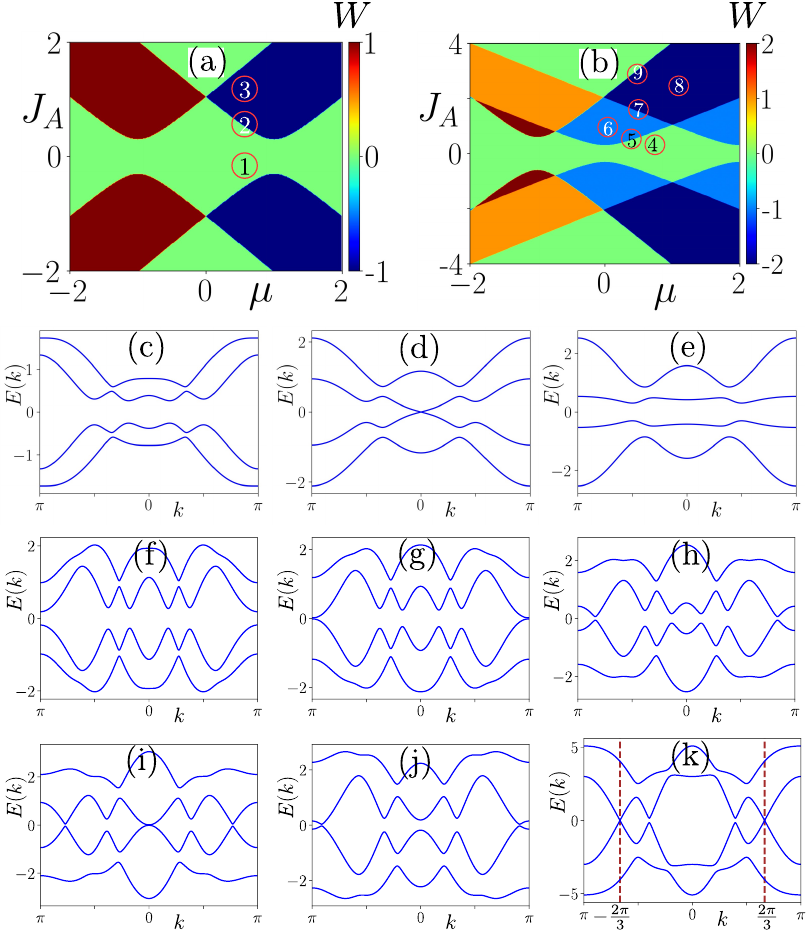}}
	\caption{Bulk band structure is illustrated for various topological phases and at the phase boundary. In panels (a),(b), we depict the phase diagram of winding number in the $\mu\!-\!J_A$ plane for the NN model ($\alpha=1,\beta=0$) and NNN ($\alpha=\beta=1$) model respectively and highlight nine points (1)-(9) corresponding each topological phase and on the phase boundary. 
	Panels (c)-(e) correspond to the points (1)-(3), respectively as marked in panel (a) whereas panels (f)-(k) correspond to points (4)-(9) as highlighted in panel (b). Model parameters for panel 
	(c)-(e) are chosen as ($\mu/t_h,J_A/t_h$)=(0.5, -0.2), (0.5, 0.583), (0.5, 1), respectively. In panels (f)-(k), we choose the following values of model parameters ($\mu/t_h,J_A/t_h$): (0.7,0.5), (0.5, 0.583), (0.1, 1.0), (0.55, 1.48), (1.2, 2.5), (0.55,3.15) respectively. We fix $\Delta/t_{h}=0.3$ and 
$\lambda_{R}/t_{h}=0.5$ for all the plots.}
	\label{fig:bulk_band}
\end{figure}

In this section, we obtain the explicit expression for bulk band structure corresponding to the Hamiltonian mentioned in Eq.\,(1) of the main text and comprehensively study the bulk bands in momentum space for various topological phases including both the nearest neighbor (NN) and next nearest neighbor (NNN) model. After diagonalizing the Hamiltonian, we obtain the following expression for bulk bands,
\begin{equation}
	E(k)= \pm \sqrt{J_k^2 + \xi_k^2 + \lambda_k^2 + \Delta^2 \pm 2 \sqrt{J_k^2 (\xi_k^2 + \Delta^2) + \xi_k^2 \lambda_k^2}}\ ,
\end{equation} 
where, $\xi_k = t_h(\alpha \cos k + \beta \cos 2k)-\mu$, $J_k = J_A\cos k$, and  $\lambda_k=\lambda_{R}(\alpha \sin k + \beta \sin 2k)$. Note that for NN model, $\alpha=1, \beta=0$ and for NNN model, $\alpha=\beta=1$ have been assumed.
We illustrate the behavior of the bulk bands in various topological phases of our model in Fig.\,\ref{fig:bulk_band}. In Fig.\,\ref{fig:bulk_band}\,(a),(b), we depict the phase diagram of winding number, $W$, in the $\mu-J_A$ plane for both NN and NNN model, respectively and highlight nine points- (1)-(9), for which we present the bulk band structures in Fig.\,\ref{fig:bulk_band}(c)-(k). We also depict the bulk band structure at the phase boundary of the topological phases where bulk gap closes. Specifically, Fig.\,\ref{fig:bulk_band}(c),(d),(e) represent the band structure for the NN model corresponding to the points (1),(2),(3) respectively, as highlighted in Fig.\,\ref{fig:bulk_band}(a), whereas Fig.\,\ref{fig:bulk_band}(f)-(k) represent the band structure for NNN model corresponding to the points, (4)-(9), as marked in Fig.\,\ref{fig:bulk_band}(b). 
Interestingly, the various phases with definite values of $W$, the bulk is always gapped whereas for the points on the phase boundary, (2), (5), (7), and (9), the valence and conduction band touch and leads to gapless state. This feature of vanishing bulk gap in the parameter space is also demonstrated in Fig.\,\ref{fig:S2}. Importantly, from the analysis of gapless lines [Eq.~(\ref{eq:sup_condition})] in Sec.~\ref{subsec:ana_gap}, we find the critical momentum values where the bulk gap closes as, $k_c=0, \pm \pi, \pm \cos^{-1}(-\frac{\alpha}{2\beta})$. This also matches with band structure in the gapless phase for both the NN and NNN case. In particular, the critical momenta for the NNN model, are $k_c=0,\pm 1, \pm \frac{2\pi}{3}$ and $k_c=\pm \frac{2\pi}{3}$ is also highlighted in 
Fig.\,\ref{fig:bulk_band}(k) by the red dashed lines.

\vspace{-0.3cm}
\section{Allowed range of the next nearest neighbor~(NNN) hopping strength}
Throughout our study, we consider NNN hopping strength to be equal to the NN hopping to 
avoid complexity in numerical and analytical calculations.
Although, practically NNN hopping strength is in general smaller than the NN one \ie $\beta < \alpha$. In this section, we explore the flexibility or constraint on the choice of $\beta$. In subsection~\ref{subsec:ana_gap}, referring to the situation number 3 that represents the extra gapless line which appears in the NNN model: $\cos{k_{*}}=-\frac{\alpha}{2 \beta}$. This makes a constraint on the choice of $\beta$: $|\beta|\geq |\alpha /2|$. Hence, to obtain $|W|=2$ in addition to $|W|=1$ in NNN model, one has to satisfy $|\beta| > |\alpha/2|$. Here, considering $\beta=0.6 \alpha$, we repeat 
the phase diagram for the number of zero modes $N_{0}$ employing OBC~[see Fig.~\ref{fig:beta}(a)] and winding number W~[see Fig.~\ref{fig:beta}(b)]. We obtain similar qualitative results as discussed for $\beta=\alpha$.

\begin{figure}[H]
\vspace{-0.2cm}
	\centering
	\subfigure{\includegraphics[width=1.0\textwidth]{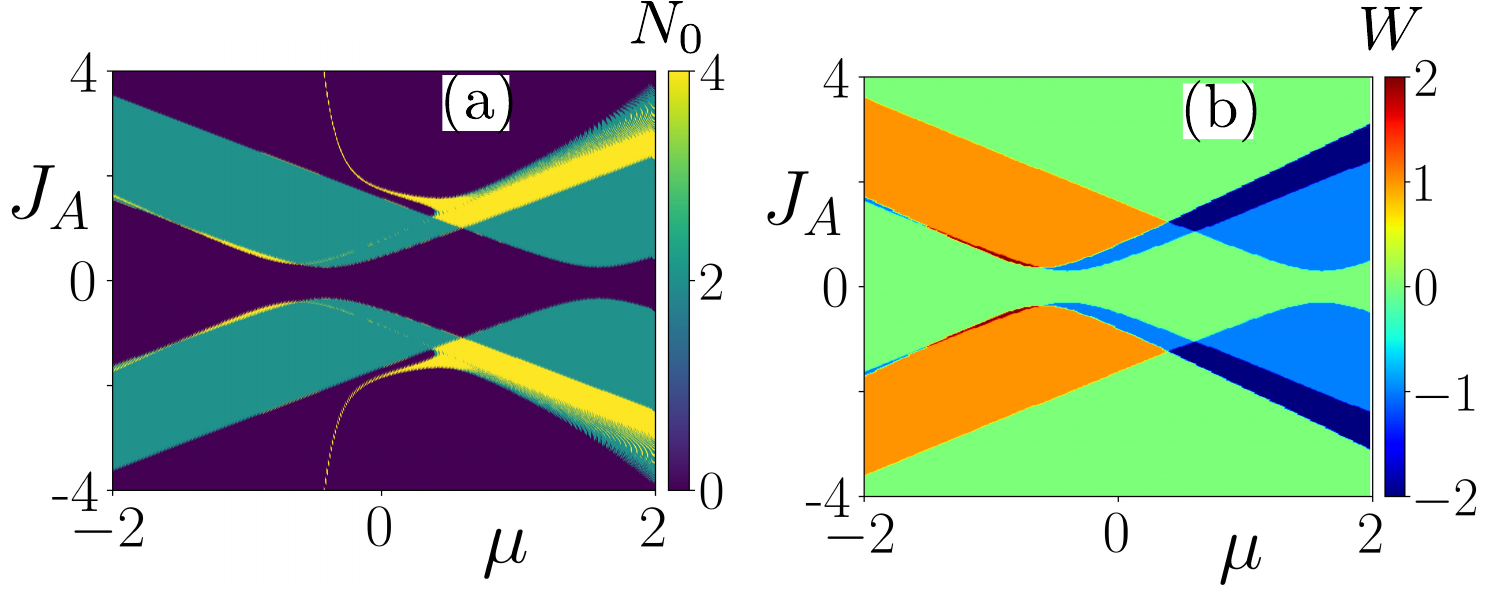}}
	\caption{We repeat the phase diagram described in Fig.~2(c) and 2(d) of main text 
considering weaker NNN hopping strength~($\beta=0.6 \alpha$). Rest of the model parameters remain same as mentioned in Fig.~2(c) and 2(d), respectively in the main text.}
	\label{fig:beta}
\end{figure}

\section{Analytics on transport and shot noise in T-junction setup}~\label{sec:formalism}
	Here, following the Refs.~\cite{BLANTER20001,Anantram1996,HaimPRL2015,Haim_PRB2015}, we present the explicit expressions for the current and their cross correlations in the T-junction set up at finite temperature employing scattering matrix formalism. In a typical normal metal-superconductor (SC) junction, an electron, incident on the interface from the leads, can reflect back as an electron (normal reflection), can reflect back as a hole with a Cooper pair transmitted inside the SC (Andreev reflection) or can transmit inside the SC as electron or hole like Bogoliubov quasiparticles. However, 
if the energy of the incident electron is less than the superconducting gap, $\Delta$ (subgap regime), then due to absence of single-particle excitations inside SC, all the scattering processes can be described in terms of normal and Andreev reflection solely. Thus the scattering matrix in this set up can be represented by only the reflection matrix containing normal and Andreev reflection amplitudes.

	Let us denote the incoming electron channels into the scattering region of the `T'-junction, originating from left and right normal leads by $\hat{a}^e_L$ and $\hat{a}^e_R$ respectively. If there are `$2M$' transverse modes in each lead (counting both spin up and down degrees of freedom), then we can compactly write the incoming channels as $\hat{a}^e_i$ where $i=1,2,...,2M$ ($i=2M+1,...,4M$) represents the channels in the right (left) lead. Similarly, incoming channels for holes can be written as $\hat{a}^h_i$ where $i=1,2,...,4M$. Outgoing channels for electron and hole in the $i^{\rm{th}}$ channel in the leads is denoted as $\hat{b}^e_i$ and $\hat{b}^h_i$ with $i=1,2,...,4M$. Now, the outgoing and incoming channels are related via the scattering matrix (reflection matrix, $r_{\rm{tot}}$, 
	in this case) as, 
	\begin{eqnarray}
		\begin{pmatrix}
			\hat{b}^e \\ \hat{b}^h
		\end{pmatrix}=r_{\rm{tot}}
		\begin{pmatrix}
			\hat{a}^e \\ \hat{a}^h
		\end{pmatrix},\mathrm{where \,\,\, } 
		r_{\rm{tot}}=\begin{pmatrix}
			r^{ee}&r^{eh}\\
			r^{he}&r^{hh}
		\end{pmatrix}\ , \label{eq:rtot}
	\end{eqnarray} or equivalently, 
	$\hat{b}^\alpha_i=\sum_{\beta,j} r^{\alpha \beta}_{ij} \hat{a}^\beta_j$
	with $\alpha,\beta={e,h}$. Here, each block of `$r_{\rm{tot}}$'  represents a $4M\times4M$ matrix. Moreover, the matrix element $r_{ij}^{\alpha\beta}$ represents the scattering amplitude of a 
	$\beta$-type particle in channel $j$ to be reflected into channel $i$ as an $\alpha$-type particle where $\alpha,\beta\in \{e,h\}$. We follow the convention where the Greek indices $\alpha,\beta,...$ represent electron-hole whereas the Latin indices like $i,j,...$ are for transverse channels in the leads. 
	
	Then, to obtain the current in our `T'-junction set up, we start with the current operator for electrons and holes in $i^{\rm{th}}$ channel in the lead as~\cite{BLANTER20001,Anantram1996},
	\begin{equation}
		\hat{I}^\A_i (x,t)= \mathrm{sgn}(\A) \frac{e\hbar}{2m i} \left(\hat{\Psi}^{\dagger}_\A(i,x)\frac{d\hat{\Psi}_\A(i,x)}{dx}- \frac{d\hat{\Psi}^\dagger_\A(i,x)}{dx}\hat{\Psi}_\A(i,x)\right)\ , \label{Eq.current_operator}
	\end{equation}
	where, the field operators $\hat{\Psi}_\A(i,x)$ and $\hat{\Psi}^\dagger_\A(i,x)$ are given by, 
	\begin{equation}
		\Psi_\A(i,x)=\int \,dE\, e^{-iEt/\hbar} \frac{\chi _i(x)}{(2\pi\hbar v_i)^{1/2}} \left[\hat{a}^\A_i e^{ik_ix} + \hat{b}^\A_i e^{-ik_ix}\right]\ , \label{Eq.chi}
	\end{equation} and 
	\begin{equation}
		\Psi^\dagger_\A(i,x)=\int \,dE\, e^{iEt/\hbar} \frac{\chi^* _i(x)}{(2\pi\hbar v_i)^{1/2}} \left[\hat{a}^{\dagger\A}_i e^{-ik_ix} + \hat{b}^{\dagger\A}_i e^{ik_ix}\right]\ ,  \label{Eq.chi*}
	\end{equation} 
	where, $\chi_i(x)$ is the transverse wave function in the $i^{\rm{th}}$ channel and $\text{sgn}(\alpha)=1 (-1)$ for $\alpha=e (h)$. Substituting Eq.~\eqref{Eq.chi} and Eq.~\eqref{Eq.chi*} into 
	Eq.~\eqref{Eq.current_operator} and employing orthonormality condition of the transverse wave functions, we obtain the following expression \cite{BLANTER20001,Anantram1996}, 
	
	\begin{equation}
		\hat{I}^\A_i(t) = \sgn (\A)\frac{e}{h} \int dE \,dE^\prime e^{i(E-E^\prime)t/\hbar} \left[\hat{a}^{\dagger\A}_i(E) \hat{a}^{\A}_i(E^\prime) - \hat{b}^{\dagger\A}_i(E) \hat{b}^\A_i(E^\prime)\right]\ .
	\end{equation}
	
	Afterwards, the total current in the normal leads due to both electron and hole can be written as, 
	\begin{eqnarray}
		\hat{I}(t) &=& \frac{e}{h} \int dE \, dE^\prime   e^{i(E-E^\prime)t/\hbar}\,  \sum_{i,\A}\sgn (\A) \left[\hat{a}^{\dagger\A}_i(E) \hat{a}^{\A}_i(E^\prime) - \hat{b}^{\dagger\A}_i(E) \hat{b}^\A_i(E^\prime)\right]\ , \non \\
		&=& \frac{e}{h} \int dE \, dE^\prime   e^{i(E-E^\prime)t/\hbar}\, \sum_{i,\A} \sgn (\A) \left[\hat{a}^{\dagger\A}_i(E) \hat{a}^{\A}_i(E^\prime) - \sum_{\G,\D}\sum_{j,k} (r^{\A\G}_{ij})^*(r^{\A\D}_{ik})\hat{a}^{\dagger\G}_j(E) \hat{a}^\D_k(E^\prime)\right]\ , \non \\
		&=& \frac{e}{h} \int dE \, dE^\prime  e^{i(E-E^\prime)t/\hbar}\sum_{\G,\D}\sum_{j,k} \sum_{i,\A}\, \sgn (\A) \left[\D_{ij}\D_{jk}\D_{\A\G}\D_{\A\D} -  (r^{\A\G}_{ij})^*(r^{\A\D}_{ik})\right]\hat{a}^{\dagger\G}_j(E) \hat{a}^\D_k(E^\prime)\ , \non \\	
		&=&\frac{e}{h} \int dE \, dE^\prime  e^{i(E-E^\prime)t/\hbar}\sum_{\G,\D}\sum_{j,k} \sum_{i,\A}\, \sgn (\A) \left[A^{\G\D}_{jk}(i,\A)\right]\hat{a}^{\dagger\G}_j(E) \hat{a}^\D_k(E^\prime)\ . \label{Eq.current_op2}
	\end{eqnarray}
	where, $ A_{jk}^{\gamma\delta}(i,\alpha;E)=\delta_{ij} \delta_{ik} \delta_{\alpha \gamma} \delta_{\alpha \delta}-(r_{ij}^{\alpha\gamma})^{*} r_{ik}^{\alpha\delta}$. We obtain the average current by 
	taking the expectation value of Eq.~\eqref{Eq.current_op2} as, 
	\begin{equation}
		I=\langle \hat{I}(t)\rangle=\frac{e}{h}\sum_{k,i=1,..,4M} \sum_{\alpha,\G \in\{e,h\}} \text{sgn}(\alpha) \int_{0}^{\infty} dE \, A_{kk}^{\G\G}(i,\alpha;E) \, f_{\G}(E) \label{eq:current_general}\ ,
	\end{equation}
	Here, we have used the relation: $\langle \hat{a}^{\dagger\G}_j(E) \hat{a}^\D_k(E^\prime)\rangle = \D_{jk}\D_{\G,\D} \D(E-E^\prime) f_\G(E)$ with $f_{e}(E)=1-f_{h}(-E)=1/\{1+\exp[(E-eV)/k_{B}T]\}$.   
	Also, $f_e(E)~(f_h(E))$ denotes the Fermi-Dirac distribution function of electrons (holes) in the reservoirs (attached to the leads) with applied bias voltage $V$.
	
	Afterwards, we turn our attention to compute the current-current correlations, also called shot noise, in this `T'-junction set up. Current-current correlations between the leads $i$ and $j$ can be 
	defined as~\cite{BLANTER20001,Anantram1996}, 
	\begin{equation}
		P_{ij}(t-t^\prime) = \frac{1}{2}\langle \Delta \hat{I}_i (t) \Delta \hat{I}_j(t^\prime) + \Delta \hat{I}_j (t^\prime) \Delta \hat{I}_i(t) \rangle \mathrm{,\,\,\,\,where\,\,} \Delta \hat{I}_i(t) = \hat{I}_i(t)  - 
		\langle \hat{I}_i(t)\rangle \label{Eq.Noise_defn}
	\end{equation} 
	Performing Fourier transformation of Eq.~\eqref{Eq.Noise_defn} from time to frequency domain, we obtain the current correlations between leads $i$ and $j$ as,
	\begin{equation}
		2\pi P_{ij}(\omega) \delta (\omega + \omega^\prime)=  \langle \Delta \hat{I}_i (\omega) \Delta \hat{I}_j(\omega^\prime) + \Delta \hat{I}_j(\omega^\prime) \Delta \hat{I}_i (\omega) \rangle \ , 
		\label{Eq.Noise_defn_omega}
	\end{equation} 
	where, $ \Delta \hat{I}_i(\omega) = \hat{I}_i(\omega)  - \langle \hat{I}_i(\omega)\rangle$. One can obtain $\hat{I}_i(\omega)$ by performing Fourier transformation of Eq.~\eqref{Eq.current_op2} as, 
	\begin{eqnarray}
		\hat{I}_i(\omega) &=& \int \, dt \, e^{i\omega t} \hat{I}_i(t) = e \int \, dE \sum_{\G,\D} \sum_\A \sum_{k,l}\sgn(\A) \hat{a}^{\G\dagger}_k(\omega) A^{\G\D}_{kl}(i,\A,E)\hat{a}^\D_l (E+\hbar \omega)\ , \label{Eq.Iomega}
	\end{eqnarray}
	Then, substituting Eq.~\eqref{Eq.Iomega} into the right hand side of the Eq.~\eqref{Eq.Noise_defn_omega} and simplifying further the expression, we obtain, 
	\begin{eqnarray}
		\langle \Delta \hat{I}_i (\omega) \Delta \hat{I}_j(\omega^\prime) +  \Delta \hat{I}_j(\omega^\prime) \Delta \hat{I}_i (\omega) \rangle\! = \!\frac{e^2}{\hbar}\! \int dE \, &\delta& (\omega + \omega^\prime)  \sum_{k,l} \!\sum_{\G,\D,\A,\B}\!\! \sgn(\A)\sgn(\B) A^{\G\D}_{kl}(i,\A,E) A^{\D\G}_{lk}(j,\B,E\!+\hbar \omega)  \non \\ &\times& f_\A(E)(1-f_\B(E\!+\!\hbar \omega))\ .  \label{Eq.IwIw'}
	\end{eqnarray}
	
	Comparing Eq.~\eqref{Eq.Noise_defn_omega} and Eq.~\eqref{Eq.IwIw'}, we obtain the expression for zero frequency current correlations between $i^{\rm{th}}$ and $j^{\rm{th}}$ channel, 
	$P_{ij}(\omega=0)$ as,
	\begin{equation}
		P_{ij}(\omega=0) = \!\frac{e^2}{h}\! \int dE \,  \sum_{k,l} \!\sum_{\G,\D,\A,\B}\!\! \sgn(\A)\sgn(\B) A^{\G\D}_{kl}(i,\A,E) A^{\D\G}_{lk}(j,\B,E) f_\A(E)(1-f_\B(E)) \ .
	\end{equation}  
	
	Since, our aim is to compute the \textit{cross}-current correlations in the `T'-junction set up, we choose $i\in R$ and $j\in L$, which leads to the following expression of cross-current correlations, $P_{RL}$, at zero frequency as~\cite{BLANTER20001,Anantram1996,HaimPRL2015,Haim_PRB2015},
	
\begin{equation}
	P_{RL}= \frac{e^{2}}{h} \sum_{i\in R,j\in L}\,\sum_{k,l=1,..,4M} \sum_{\alpha,\beta,\gamma,\delta \in\{e,h\}} \sgn(\alpha) \, \sgn(\beta) \int_{0}^{\infty} dE   A_{kl}^{\gamma\delta}(i,\alpha;E) \, A_{lk}^{\delta \gamma}(j,\beta;E) \, f_{\gamma}(E)\left[ 1 - f_{\delta}(E)\right] \label{eq:shot_noise_general}\ ,
\end{equation}
	
	Therefore, Eq.~\eqref{eq:current_general} and Eq.~\eqref{eq:shot_noise_general} serve as the general expression to compute current and shot noise in `T'-junction heterostructure.
	In terms of the blocks of $r_{\rm{tot}}$ matrix which is more useful for computational purpose, Eq.~\eqref{eq:current_general} and Eq.~\eqref{eq:shot_noise_general} can be written as,
	\begin{eqnarray}
		I&=&\frac{2e}{h}\int_{0}^{\infty} dE \,\, \operatorname{Tr}\left[r^{he} r^{he \dagger}\right]  f_{e}(E) \label{eq:current_finite}\ ,\\
		P_{RL}&=&\frac{e^{2}}{h}\!\! \sum_{i\in R, j\in L} \int_{0}^\infty dE \sum_{\G,\D} \Pc^{\G\D}_{ij}(E) f_{\G}(E)(1-f_\D(E))\ , \label{eq:noise_finite}
	\end{eqnarray}
	where, $\Pc^{\G\D}_{ij}(E) = \sum_{k,l=1,...,4M} \sum_{\A,\B} \sgn(\A)\sgn(\B) A^{\G\D}_{kl}(i,\A,E)A^{\D\G}_{lk}(j,\B,E)$.
	
	In the following, we drop writing the explicit energy dependence in $A^{\D\G}_{kl}(i,\A,E)$ and $f_\G(E)$ for the sake of clarity. Then we derive the corresponding expressions for $\Pc^{\G\D}_{ij}(E)$ 
	considering $\G,\D\in{e,h}$.
	
	$\bullet \,\,\G=e, \D=e: $
	\begin{eqnarray*}
		\Pc^{ee}_{ij}(E) &=& \sum_{k,l}\sum_{\A,\B} \sgn(\A)\, \sgn(\B) \,A^{ee}_{kl}(i,\A)A^{ee}_{lk}(j,\B) \\
		&=& \sum_{k,l} A^{ee}_{kl}(i,e)A^{ee}_{lk}(j,e) - A^{ee}_{kl}(i,e)A^{ee}_{lk}(j,h) - A^{ee}_{kl}(i,h)A^{ee}_{lk}(j,e) + A^{ee}_{kl}(i,h)A^{ee}_{lk}(j,h) \\ 
		&=&\sum_{kl} [\delta_{ik}\delta_{il}-(r^{ee}_{ik})^* r^{ee}_{il}][\delta_{jl}\delta_{jk}-(r^{ee}_{jl})^* r^{ee}_{jk}]  ~~- [\delta_{ik}\delta_{il}-(r^{ee}_{ik})^* r^{ee}_{il}][-(r^{he}_{jl})^* r^{he}_{jk}]  \\ &&-[-(r^{he}_{ik})^*r^{he}_{il}][\delta_{jl}\delta_{jk}-(r^{ee}_{jl})^* r^{ee}_{jk}] + [-(r^{he}_{ik})^*r^{he}_{il}][-(r^{he}_{jl})^* r^{he}_{jk}] \\
		&=& [\delta_{ij}\delta_{ji} - (r^{ee}_{ij})^* r^{ee}_{ij} - (r^{ee}_{ji})^* r^{ee}_{ji} + ( r^{ee}r^{ee}{^\dagger} )_{ji} (r^{ee}r^{ee}{^\dagger})_{ij}] + \\
		&&[-(r^{he}_{ji})^* r^{he}_{ji} + (r^{he}r^{ee}{^\dagger} )_{ji}( r^{ee}r^{he}{^\dagger} )_{ij}] ~ + [-(r^{eh}_{ij})^* r^{eh}_{ij} + (r^{he}r^{ee}{^\dagger} )_{ji}( r^{ee}r^{he}{^\dagger} )_{ij}] \\ 
		&& + ( r^{he}r^{he}{^\dagger} )_{ji} ( r^{he}r^{he}{^\dagger} )_{ij} \\
		&=& [-|r^{ee}_{ij}|^2 - |r^{ee}_{ji}|^2 + |\mc{R}^{ee}_{ij}|^2] + ~~ [|r^{he}_{ji}|^2 - |\mc{R}^{eh}_{ji}|^2] + [|r^{he}_{ij}|^2 - |\mc{R}^{eh}_{ij}|^2] + |\mc{R}^{hh}_{ij}|^2 \\
		&=& |r^{he}_{ij}|^2 + |r^{he}_{ji}|^2 - |r^{ee}_{ij}|^2 - |r^{ee}_{ji}|^2 + |\mc{R}^{ee}_{ij}|^2 - |\mc{R}^{eh}_{ij}|^2 - |\mc{R}^{he}_{ij}|^2 + |\mc{R}^{hh}_{ij}|^2\ ,
	\end{eqnarray*}
	
	$\bullet \G=e, \D=h :$
	\begin{eqnarray*}
		\Pc^{eh}_{ij}(E) &=& \sum_{k,l}\sum_{\A,\B} \sgn(\A)\, \sgn(\B) \,A^{eh}_{kl}(i,\A)A^{he}_{lk}(j,\B) \\
		&=& \sum_{k,l} A^{eh}_{kl}(i,e)A^{he}_{lk}(j,e) - A^{eh}_{kl}(i,e)A^{he}_{lk}(j,h) - A^{eh}_{kl}(i,h)A^{he}_{lk}(j,e) + A^{eh}_{kl}(i,h)A^{he}_{lk}(j,h) \\ 
		&=&\sum_{kl} [(r^{ee}_{ik})^* r^{eh}_{il}][(r^{eh}_{jl})^* r^{ee}_{jk}]  ~~- [(r^{ee}_{ik})^* r^{eh}_{il}][(r^{hh}_{jl})^* r^{he}_{jk}]  \\ &&-[(r^{he}_{ik})^*r^{hh}_{il}][(r^{eh}_{jl})^* r^{ee}_{jk}] + [(r^{he}_{ik})^*r^{hh}_{il}][(r^{hh}_{jl})^* r^{he}_{jk}] \\
		&=& (r^{ee}r^{ee}{^\dagger} )_{ji} (r^{eh}r^{eh}{^\dagger})_{ij} - (r^{he}r^{ee}{^\dagger} )_{ji} (r^{eh}r^{hh}{^\dagger} )_{ij} - (r^{ee}r^{he}{^\dagger} )_{ji} (r^{hh}r^{eh}{^\dagger} )_{ij} + (r^{he}r^{he}{^\dagger} )_{ji} (r^{hh}r^{hh}{^\dagger} )_{ij}\\
		&=& [ \delta_{ij}\mc{R}^{ee}_{ji}-|\mc{R}^{ee}_{ij}|^2 + |\mc{R}^{eh}_{ij}|^2 + |\mc{R}^{he}_{ij}|^2 + \delta_{ij}\mc{R}^{hh}_{ji} -|\mc{R}^{hh}_{ij}|^2] \\
		&=&  |\mc{R}^{eh}_{ij}|^2 + |\mc{R}^{he}_{ij}|^2  -|\mc{R}^{ee}_{ij}|^2  -|\mc{R}^{hh}_{ij}|^2\ ,
	\end{eqnarray*}

	$\bullet \G=h, \D=e :$
	\begin{eqnarray*}
		\Pc^{he}_{ij}(E) &=& \sum_{k,l}\sum_{\A,\B} \sgn(\A)\, \sgn(\B) \,A^{he}_{kl}(i,\A)A^{eh}_{lk}(j,\B) \\
		&=& \sum_{k,l}\sum_{\A,\B} \sgn(\B)\, \sgn(\A) \,A^{he}_{lk}(i,\B)A^{eh}_{kl}(j,\A) \\
		&=& \sum_{k,l}\sum_{\A,\B} \sgn(\A)\, \sgn(\B) \,A^{eh}_{kl}(j,\A)A^{he}_{lk}(i,\B) \\
		&=& \Pc^{eh}_{ji}(E)\ ,
	\end{eqnarray*}
	
	$\bullet \,\,\G=h, \D=h: $
	\begin{eqnarray*}
		\Pc^{hh}_{ij}(E) &=& \sum_{k,l}\sum_{\A,\B} \sgn(\A)\, \sgn(\B) \,A^{hh}_{kl}(i,\A)A^{hh}_{lk}(j,\B) \\
		&=& \sum_{k,l} A^{hh}_{kl}(i,e)A^{hh}_{lk}(j,e) - A^{hh}_{kl}(i,e)A^{hh}_{lk}(j,h) - A^{hh}_{kl}(i,h)A^{hh}_{lk}(j,e) + A^{hh}_{kl}(i,h)A^{hh}_{lk}(j,h) \\ 
		&=&\sum_{kl} [-(r^{eh}_{ik})^* r^{eh}_{il}][-(r^{eh}_{jl})^* r^{eh}_{jk}]  ~~- [-(r^{eh}_{ik})^* r^{eh}_{il}][\delta_{jl}\delta_{jk}-(r^{hh}_{jl})^* r^{hh}_{jk}]  \\ &&-[\delta_{ik}\delta_{il}-(r^{hh}_{ik})^*r^{hh}_{il}][-(r^{eh}_{jl})^* r^{eh}_{jk}] + [\delta_{ik}\delta_{il}-(r^{hh}_{ik})^*r^{hh}_{il}][\delta_{jl}\delta_{jk}-(r^{hh}_{jl})^* r^{hh}_{jk}] \\
		&=& [|r^{eh}_{ij}|^2 + |r^{eh}_{ji}|^2 - |r^{hh}_{ij}|^2 - |r^{hh}_{ji}|^2 + |\mc{R}^{ee}_{ij}|^2 - |\mc{R}^{eh}_{ij}|^2 - |\mc{R}^{he}_{ij}|^2 + |\mc{R}^{hh}_{ij}|^2]
	\end{eqnarray*}
	
	where, $\mc{R}^{\A\B}_{ij}= [r^{\A e} (r^{\B e})^{\dagger}]_{ij}$.  We use the fact that $\delta_{ij}=0$ if $i\in R$ and $j\in L$. We have also employed the unitarity relation of the reflection matrix, $r_{\rm{tot}}$ \ie $r_{\rm{tot}}^\dagger r_{\rm{tot}}=\mc{I}$.

	At zero temperature, Eqs.~\eqref{eq:current_finite} and \eqref{eq:noise_finite} acquire simplified expressions as~\cite{Haim_PRB2015,Haim_PRL_2015}
	\begin{eqnarray}
		I&=&\frac{2e}{h}\int_{0}^{eV} dE \hspace*{1 mm} \operatorname{Tr}\left[r^{he} r^{he \dagger}\right] \label{eq:current_zero}\ ,\\
		P_{RL}&=&\frac{e^{2}}{h}\sum_{i\in R,j\in L} \int_{0}^{eV} dE \hspace*{2 mm} {\cal{P}}_{ij}^{eh}(E) \label{eq:noise_zero}\ .
	\end{eqnarray}


	\section{Coupling matrix $W_{c}$}~\label{sec:Weidenmuller}
We use an energy independent coupling matrix $W_{c}$ in Eq.~(5) in the main text. Here, in this section we show the form of $W_{c}$ that we use in our analysis. In real space, considering the following BdG basis: $\Psi_{j}=\left\{\psi_{j \ua},\psi_{j \da}, \psi_{j \da}^{\dagger},-\psi_{j \ua}^{\dagger}  \right\}^{\mathbf{t}}$, under open boundary condition (OBC) with $N$-number of sites, the system Hamiltonian~[Eq.~(1) in main text] can be written as
\begin{eqnarray}
    H=&&\sum_{jl} \Psi_{j}^{\dagger} \left[-\mu ~\tau_{z}\sigma_{0} +\Delta~\tau_{x}\sigma_{0}\right]\delta_{jl}~\Psi_{l}+\frac{\alpha}{2}\sum_{jl}\Psi_{j}^{\dagger} \left[t_{h}~\tau_{z}\sigma_{0}-i \lambda_{R}~\tau_{z}\sigma_{y} +J_{A}~\tau_{0}\sigma_{z}\right]\delta_{j+1,l}~\Psi_{l}\non\\
    &&+\frac{\beta}{2}\sum_{jl}\Psi_{j}^{\dagger} \left[t_{h}~\tau_{z}\sigma_{0}-i \lambda_{R}~\tau_{z}\sigma_{y} +J_{A}~\tau_{0}\sigma_{z}\right]\delta_{j+2,l}~\Psi_{l}+H.c. \label{eq:real_Hamiltonian}\\
    =&& \sum_{jl}\Psi_{j}^{\dagger} (H_{{\rm{TB}}})_{jl}~\Psi_{l}\ .\non
\end{eqnarray} 
Hence, the first quantized Hamiltonian $H_{\rm{TB}}$ is a $4N\times 4N$ matrix and we diagonalize this $H_{{\rm {TB}}}$ to obtain the spectra employing OBC. The $W_{c}$ matrix in Eq.~(5) of main text connects the transverse modes in the leads with the lattice sites of the NW as shown in Fig.~\ref{fig:S3}. If $M$ number of sites of the NW are connected to the leads, then in each lead there are $M$ number of transverse modes for each spin species \ie total $2M$ number of transverse modes are participating in each of the leads. Encompassing both the spin species, particle-hole degrees of freedom and two leads, $W_{c}$ can be written as a $4N\times 8M$ matrix of the form~\cite{Haim_PRB2015}:
\begin{eqnarray}
    W_{c}=\begin{pmatrix}
        W_{e}&0\\
        0&-W_{e}^{*}
    \end{pmatrix}; ~~~ W_{e}=(W_{L}~~W_{R}) \label{eq:Wc}.
\end{eqnarray}
Here, $W_{L}$ and $W_{R}$ represents the coupling with left and right leads for particles and given by~\cite{Haim_PRB2015}
\begin{eqnarray}
    W_{L}&=&W_{R}={\mathcal{W}} \otimes \sigma_{0},\non\\
    \mathcal{W}_{nm}&=&
    \begin{cases}
      w_{m} \sin\left(\frac{\pi n m}{M+1} \right), & 1\leq n \leq M \\
      ~~~~~~~0, & M < n \leq N
    \end{cases},~~ m=1,.....,M,
\end{eqnarray}
with $\sigma_{0}$ is a $2 \times 2$ identity matrix in spin-space, and $w_{m}$ denotes a set of coupling constants for each transverse channel of the leads. We consider $M=1$ for the system hosting single Majorana pair and $M=2$ for the system hosting double Majorana pairs considering $w_{m}^{2}=0.01 ~t_{h},~\forall m \in \{1,2\}$.
\vspace{+0.3cm}

\begin{figure}
	\centering
	\subfigure{\includegraphics[width=0.28\textwidth]{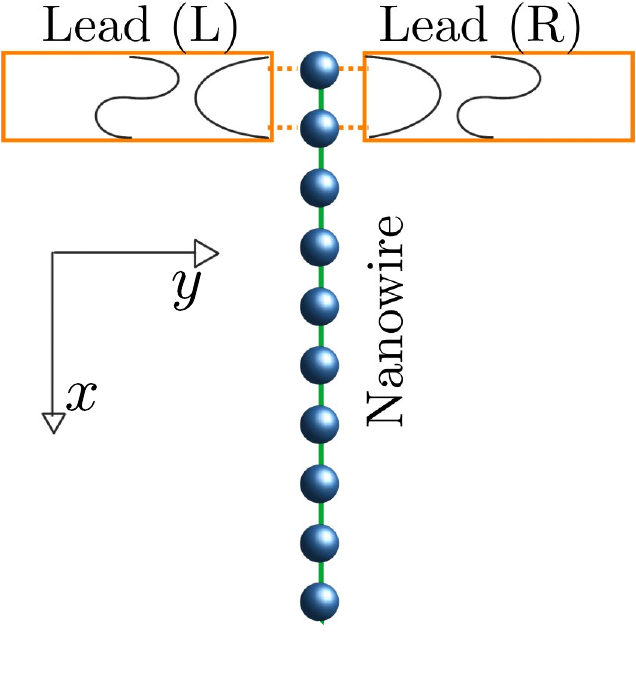}}
	\caption{Schematic illustration of the NW as tight binding model. The lattice sites are represented by blue circles and its coupling with the left and right leads enclosing transverse modes by orange dotted lines.}
	\label{fig:S3}
\end{figure}
\section{Finite size effect on ZBC with NNN hopping} \label{sec:finite_size}
\begin{figure}[h]
	\centering
	\subfigure{\includegraphics[width=0.7\textwidth]{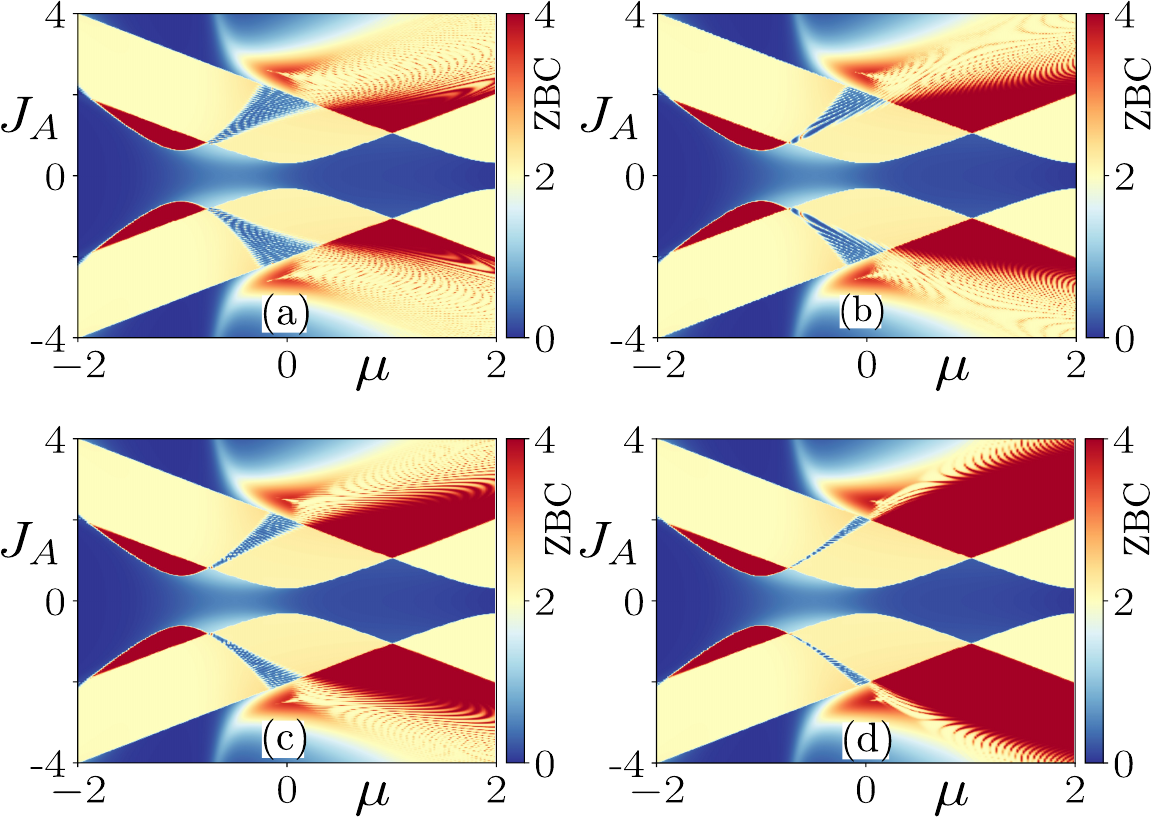}}
	\caption{We repeat Fig.~4(b) of the main text and depict here choosing four different system sizes~($N=300$,  $N=400$, $N=500$, and $N=1000$) in panels (a), (b) (c), and (d), respectively. The right top and right bottom prong-like regions correspond to the region with mixed quantization of $2 e^{2}/h$ and $4 e^{2}/h$ for ZBC. The area of such red prong-like regions is reducing and become quantized at $4 e^{2}/h$ as we increase the system size.}
	\label{fig:S31}
\end{figure}

In principle, the region with $W = -2$ in Fig.~2(d) in the main text should correspond to a 
$4e^2/h$ quantization in the zero-bias conductance (ZBC) results presented in Fig.~4(b) of main text. However, we observe a prong-like region that exhibits a mixture of $2e^2/h$ and $4e^2/h$ quantization. To investigate the underlying cause, here we perform a system-size dependence study particularly for Fig.~4(b) with NNN model. The results are shown in Fig.~\ref{fig:S31}. 
Note that, as we increase the system size, the area of this region decreases and gradually transforms into the expected $4e^2/h$ quantization (which is anticipated for $W = -2$). Hence, this indicates that the area of this prong-like region decreases as the system size increases, emphasizing that the finite system size is responsible for this effect. On the other hand, for the NN model, we choose $N=300$ lattice sites, where the results are consistent across all figures. The results remain unchanged even if we choose larger sytem size in NN case.


\section{Stability against Disorder} \label{sec:disorder}
In the main text, we have analyzed transport properties of the clean system hosting MZMs. In this section, we investigate the effect of disorder and stability of ZBC due to MZMs in both NN and 
NNN model. In order to check the disorder stability of $\frac{dI}{dV}$ against static onsite random disorder, we add a term $V(r)$ to the chemical potential $\mu$. Here, $V(r)$ is random number 
generated from a uniform distribution within range $[-V_{\rm{dis}},V_{\rm{dis}}]$ and corresponding $V_{\rm{dis}}$ denotes the disorder strength.
 \begin{figure}[H]
	\centering
	\subfigure{\includegraphics[width=1.0\textwidth]{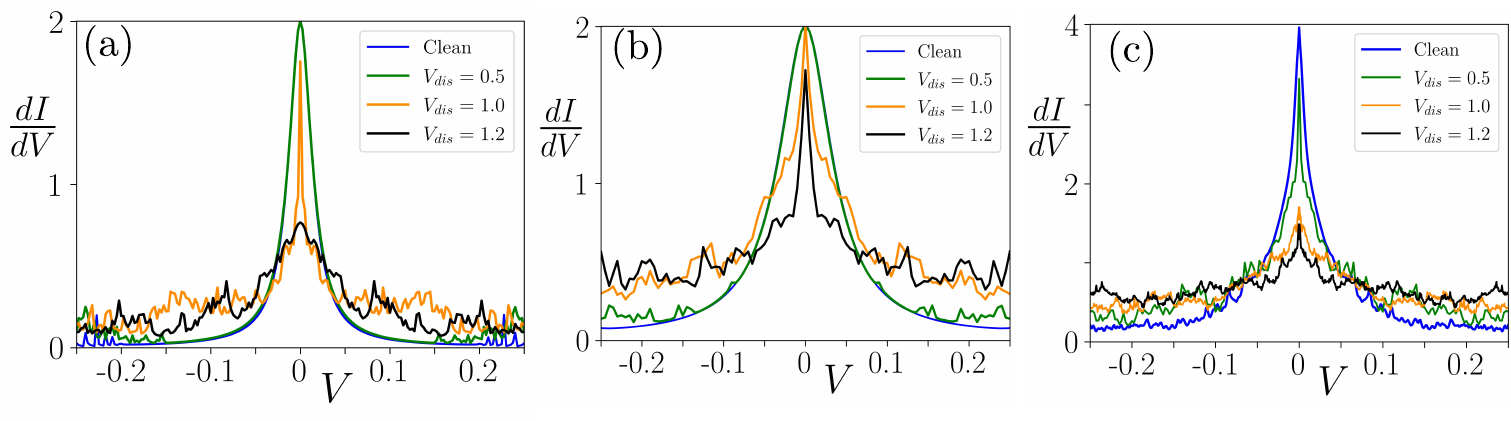}}
	\caption{(a) We depict disorder averaged $\frac{dI}{dV}$ as a function of bias voltage $V$ (at $T=0$) for MZMs with $|W|=1$ emerged in NN model~($\alpha=1,~\beta =0$). (b)~[(c)] We repeat 
	(a) for MZMs with $|W|=1$~[$|W|=2$] appeared in NNN model~($\alpha=\beta=1$). We perform disorder average considering 50 disorder configurations. All the model parameters in panel (a) 
	carry same value as mentioned in Fig.~3(a). We choose the model parameters in panel (b)~[(c)] same as mentioned in Fig.~4(a) inset I1~[Fig.~4(a)].
	}
	\label{fig:S4}
\end{figure}
In presence of disorder,  we compute $\frac{dI}{dV}$ at $T=0$ for 50 disorder configurations and then take average. For MZMs~($|W|=1$) as emerged in NN model~($\alpha=1,~\beta =0$) we compute disordered averaged $\frac{dI}{dV}$ choosing three disorder strength and depict them in Fig.~\ref{fig:S4}(a). It is evident that, for intermediate disorder strength~($V_{dis}=0.5$), $2e^{2}/h$ 
quantization remains unaffected and start losing its value for high disorder strength~($V_{dis}=1.0$). Finally for very high disorder strength~($V_{dis}=1.2$), the quantized peak height in ZBC is 
reduced significantly.  On the other hand, Fig.~\ref{fig:S4}(b)~[Fig.~\ref{fig:S4}(c)] represents disorder averaged $\frac{dI}{dV}$ for MZMs with $|W|=1$~[$|W|=2$] as emerged in NNN model~($\alpha=\beta=1$). Clearly, Fig.~\ref{fig:S4}(b) exhibits similar qualitative feature as Fig.~\ref{fig:S4}(a). However, quantitatively, the latter displays better stability against disorder than that of the former. 
For multiple MZMs, Fig.~\ref{fig:S4}(c) indicates loss of quantization of ZBC peak even at moderate disorder. Hence, ZBC of multiple MZMs~($|W|=2$) are less stable against disorder than the single 
MZM~($|W|=1$) as far as same disorder strength is concerned. Note that, low~(high) disorder strength corresponds to whether $V_{dis}$ is less~(greater) than the corresponding bulk TSC gap, and moderate disorder means $V_{dis}$ is equivalent to bulk TSC gap. This evidently is different for NN and NNN model and thus exhibits distinct stability of ZBC peak against disorder. 

\section{Transport in two terminal setup using KWANT} \label{sec:KWANT}
\begin{figure}
	\centering
	\subfigure{\includegraphics[width=0.7\textwidth]{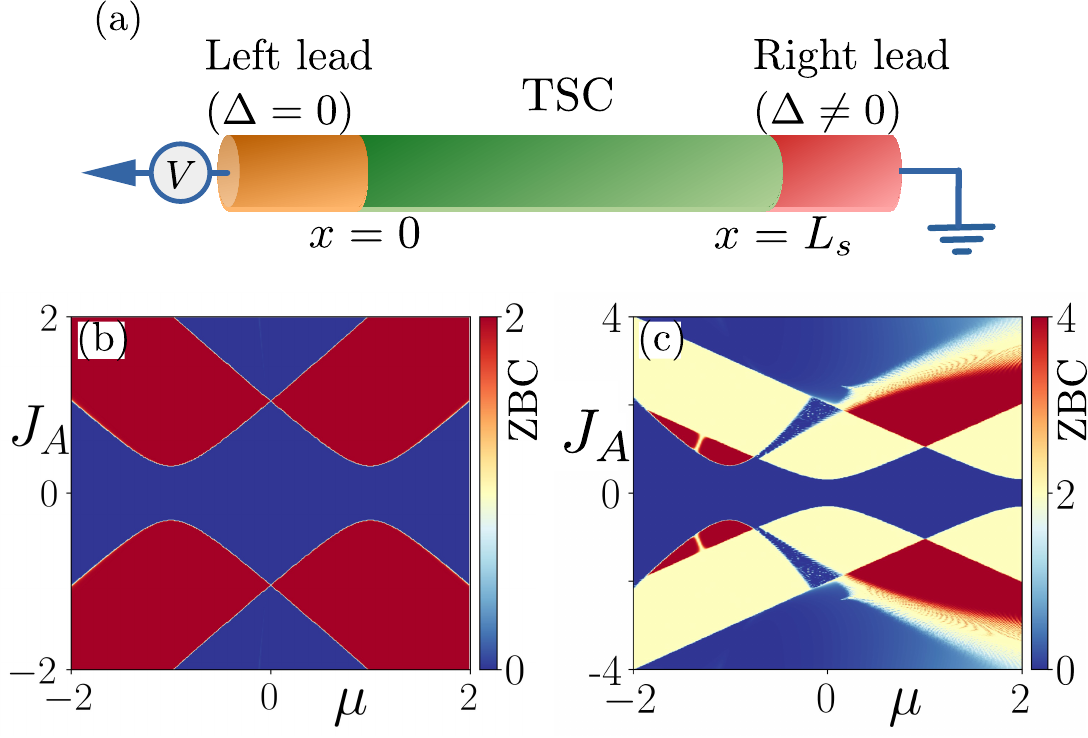}}
	\caption{(a) Schematic representation of the two terminal setup where normal metal-TSC nanowire junction is used for numerical simulation employing KWANT~\cite{KWANT_Groth_2014} to obtain 
	the ZBC . We display the ZBC (in units of $e^2/h$) in the ($\mu-J_A$) plane for the two terminal set up in (b) NN model ($\alpha=1,\beta=0$) with system size, $L_s=500$, and (c) NNN model 
	($\alpha=1,\beta=1$) with system size, $L_s=1000$. We choose all the other model parameters in the Hamiltonian as described in the main text.}
	\label{fig:S5}
\end{figure}

Here, we present our numerical simulations for differential conductance based on Blonder-Tinkham-Klapwijk (BTK) formalism considering a two terminal set up comprised of a metallic lead ($x<0$), a topological superconductor (TSC) ($0\leq x\leq L_s$), and a superconducting lead ($x\!> \!L_s$) as schematically presented in Fig.\,\ref{fig:S5}(a). In this two terminal set up, the interplay of normal and Andreev reflections give rise to all interesting phenomena. In a TSC hosting a pair of MZMs localized at the end of the SC, the ZBC becomes quantized and takes value $2e^2/h$ which has been the indirect signature (still debated) of MZMs in a transport measurement. Here, we explicitly demonstrate this fact by computing the differential conductance in such two terminal set up where the TSC 
is constructed using the Hamiltonian in Eq.~(1) of the main text. 

For this purpose, we first consider a TSC nanowire of length $L_s$ and attach leads to the two ends of it \ie at $x=0$ and $x=L_s$. The left lead ($x<0$) is chosen to be non-superconducting while the right lead ($x>L_s$) is superconducting and modeled by the same Hamiltonian used to construct the TSC. We apply a voltage bias, $V$, between the left and right lead. As mentioned earlier, we are primarily interested in the energy regime less than the superconducting gap. More specifically, here we focus only in the ZBC to obtain the signature of Majorana pairs. Therefore, the full scattering matrix is represented by only a reflection matrix containing the normal and Andreev reflection amplitudes and can be expressed as, 
\begin{equation}
	r_{\rm{NS}}=\begin{pmatrix}
		r^{ee}&r^{eh}\\
		r^{he}&r^{hh}
	\end{pmatrix}\ , \label{eq:r_NS}
\end{equation}
where, $r_{ee}$ and $r_{he}$ denote $2N_L\times2N_L$ matrices, $N_L$ being the number of transverse channels in the left lead and factor of 2 is present since both up and down spin species are counted. Here, $r_{ee}$ ($r_{he}$) represents the amplitude for normal (Andreev) reflection from the interface at $x=0$. With this setup, we employ the BTK formalism to obtain the differential conductance in a two terminal junction at zero temperature~\cite{BTK_NS1982,KWANT_Groth_2014} as 
\begin{equation}
	\frac{dI}{dV}(eV) = \frac{e^2}{h} [N_m(E) -R_{ee}(E) + R_{he}(E)]|_{E=eV}\ ,
\end{equation}

where, $R_{ee}=\mathrm{Tr} \,({r^{ee}}^\dagger r^{ee})$ and $R_{he}=\mathrm{Tr}\,({r^{he}}^\dagger r^{he})$. Also, $N_m(eV)$ denotes the number of occupied channels in the left lead for a given voltage bias $eV$. Using the Python package KWANT~\cite{KWANT_Groth_2014}, we obtain the reflection matrices, $r_{ee}$, $r_{eh}$, and $N_m$ to compute the differential conductance.

In Fig.\,\ref{fig:S5}\,(b) and Fig.\,\ref{fig:S5}\,(c), we display the results for ZBC (in units of $e^2/h$) of this setup in the $\mu$-$J_A$ plane for the NN ($\alpha=1,\beta=0$) and NNN ($\alpha=\beta=1$) model respectively [see Eq.~(1) of the main text]. In Fig.\,\ref{fig:S5}, the regions in the $\mu$-$J_A$ plane hosting MZMs exhibit its manifestation as quantized ZBC in units of $e^2/h$. For the NN model, ZBC becomes $2e^2/h$ as the model only hosts a single pair of MZMs [see Fig.\,\ref{fig:S5}(b)]. Interestingly, in  Fig.\,\ref{fig:S5}(c) for the NNN model, ZBC exhibits both $2e^2/h$ and $4e^2/h$ quantization as the model hosts both single and double pair of MZMs. In sharp contrast to this quantized peak of MZMs, AZMs do not showcase any quantized nature in the ZBC. Thus, we can safely 
infer that these are trivial modes and do not have any topological origin thus distinguishing the topological and trivial zero energy modes. We observe a one-to-one correspondence between the 
Fig.~2(b) and Fig.~2(d) of the main text with Fig.\,\ref{fig:S5}(b) and Fig.\,\ref{fig:S5}(c) respectively. Furthermore, we contrast the ZBC in this two terminal junction with ZBC obtained in case of
`T'-junction [see Fig.3(b) and Fig.\,4(b) in the main text]. Note that, for NN model, ZBC in Fig.\,\ref{fig:S5}(b) and Fig.\,3(b) (of main text) matches excellently well. Comparing Fig.\,\ref{fig:S5}(c) 
and Fig.\,4(b) (of main text), we find that for NNN model, ZBC for AZMs is more suppressed and almost close to zero in two terminal geometry compared to the ZBC in `T'-junction setup. In the latter, 
AZMs exhibit finite non-quantized value in ZBC [Fig.\,4(b) of the main text]. The reason can be multiple Andreev reflection due to the `T'-shaped region between the two leads and TSC nanowire. 
This is almost absent due to AZMs in two terminal setup. To this end, we also note that for $\mu \lesssim 0$ and $J_A \sim \pm 2$, quantization behavior of ZBC is not observed. This feature can also 
be noticed in Fig.\,4(b) of the main text. The existence of small bulk gap [see Fig.\,\ref{fig:S2}(b)] value may be the origin of this intricate behavior.

\end{onecolumngrid}
\end{document}